\documentstyle[12pt,psfig]{article}
\newcommand{\NP}[1]{ Nucl.\ Phys.\ {\bf #1}}
\newcommand{\ZP}[1]{ Z.\ Phys.\ {\bf #1}}

\newcommand{\PL}[1]{ Phys.\ Lett.\ {\bf #1}}

\newcommand{\PR}[1]{Phys.\ Rev.\ {\bf #1}}
\newcommand{\PRL}[1]{ Phys.\ Rev.\ Lett.\ {\bf #1}}

\newcommand{\mpi}{m_\pi^2}
\newcommand{\mk}{m_K^2}
\newcommand{\me}{m_\eta^2}
\newcommand{\sq}{\mpi + \mk}
\newcommand{\fpi}{f_\pi^2}
\newcommand{\fpc}{f_\pi^4}
\newcommand{\fk}{f_K^2}
\newcommand{\fe}{f_\eta^2}
\newcommand{\fkc}{f_K^4}

\newcommand{\nn}{\nonumber}
\newcommand{\spa}{\quad\quad\quad}
\newcommand{\La}{{\cal L}}
\newcommand{\oK}{\overline{K}}
\newcommand{\Kma}{K^+}
\newcommand{\Kme}{K^-}
\newcommand{\Opd}{{\cal O}(p^2)}
\newcommand{\Opc}{{\cal O}(p^4)}
\newcommand{\Ima}{{\rm Im}\,}
\newcommand{\Rea}{{\rm Re}\,}
\begin{document}

\thispagestyle{empty}

\hfill  SLAC-PUB-7787
\vspace*{2cm}
\begin{center}
{\LARGE{\bf Meson-Meson interaction in a \\ 
\vspace{0.2cm}
non-perturbative chiral approach}}
\end{center}

\vspace{2cm}

\begin{center}
{\Large J. A. Oller$^{a,1}$, E. Oset$^{a,2}$, J. R. Pel\'aez$^{b,3}$}

\vspace{1.cm}

$^a$ {\it Departamento de F\'{\i}sica Te\'orica and IFIC

Centro Mixto Universidad de Valencia-CSIC,

46100 Burjassot (Valencia), Spain}

\vskip .5cm

$^b$ {\it Stanford Linear Accelerator Center\\
 Stanford University, 
Stanford, California 94309}

\end{center}

\vskip 1.5 cm

\begin{abstract}
 A non-perturbative method which combines constraints from chiral
 symmetry breaking and coupled channel unitarity is used to
 describe the meson-meson interaction up to about 1.2 GeV. The
 approach uses the  $\Opd$  and $\Opc$ chiral Lagrangians. 
The seven free parameters of the 
$\Opc$ Lagrangian are fitted to the data. The results are in good
agreement with a vast amount of experimental analyses. The amplitudes 
develop poles in the complex plane corresponding to the $f_0,
a_0, \rho,K^*,\phi, \sigma$ and $\kappa$ resonances; the latter two, very broad.
The total and partial decay widths of the resonances are also
well reproduced. Further extensions and applications of this chiral
non-perturbative scheme are also discussed.

\end{abstract}

\footnotetext[1]{E-mail:oller@titan.ific.uv.es}
\footnotetext[2]{E-mail:oset@evalvx.ific.uv.es}
\footnotetext[3]{E-mail:pelaez@eucmax.sim.ucm.es\\
On leave of absence from Departamento de F\'{\i}sica Te\'orica, 
Universidad Complutense, 28040 Madrid, Spain}

\newpage

\section{Introduction}
The  meson-meson interaction has been the key problem to test 
Chiral Perturbation Theory ($\chi PT$), which has proved rather successful at
low energies \cite{GaLe,ChPT}. The underlying idea is that an expansion in 
powers of the meson momenta converges at sufficiently low energy, which in
practice is $\sqrt{s} \leq 500$  MeV. However, the convergence at 
higher energies 
becomes progressively worse. Even more, one of the peculiar features
of the meson-meson interaction is the presence of resonances like the
$\sigma, f_0, a_0$ in the scalar sector and the $\rho,
K^*$ or the $ \phi$ in the vector channels. These resonances will show up
in the $T$ matrix as poles that cannot be obtained 
using standard $\chi PT$. Nevertheless, the
constraints imposed by chiral symmetry breaking are rather powerful
and not restricted to the region where $\chi PT$ is  meant to converge 
\cite{Steele}.

Two independent approaches of non perturbative character have extended 
the use
of chiral Lagrangians to higher energies and have been rather successful, 
reproducing important features of the  meson-meson interaction
including several resonances. 
Although these two approaches  look in principle rather different,
they share a common feature which is the imposition
of  unitarity. One of them \cite{truongdo,IAM},  based upon the
Inverse Amplitude Method (IAM), first suggested in 
\cite{truong}, makes use of the lowest order, $\Opd$, as well as the next 
to leading order, $\Opc$, Lagrangians. Elastic unitarity is 
imposed and thus
no mixture of channels is allowed. Then, the coefficients of
the $\Opc$ Lagrangian are fitted to the data.
The absence of coupled channels has obvious limitations, but 
in spite of them,
 the IAM is able to generate dynamically the $\rho$, $K^*$ and $\sigma$ 
 resonances, and to reproduce $\pi\pi$ scattering in the (I,J)=(0,0),
(1,1), (2,0) partial waves, as well as in the (3/2,0),(1/2,1)
and (1/2,0) channels of $\pi K$ scattering. The results are very
successful up to 1 GeV in all these channels but the (0,0), where it
only yields good results up to 700 MeV. 
The limitations of this single
channel approach become evident, for instance, in  the $f_0(980)$ 
and $a_0(890)$
resonances (J=0 and I=0 and 1, respectively) which do not 
appear as poles in the
T matrix. The method also has a pathological behavior close to the
T matrix zeros \cite{Pennington}. 

The second approach dealt with the J=0 sector alone \cite{olset}. The 
input
consists of the $\Opd$ Lagrangian, which is used as the source of a
potential between mesons. This potential enters in a set
of coupled channel Lippmann-Schwinger (LS) equations  (actually closer
to Bethe-Salpeter equations, since relativistic propagators are used) 
which leads to the scattering
matrix. The method imposes unitarity in coupled
channels; hence it yields inelasticities when inelastic 
channels open up. Amazingly, the approach has only one free parameter,
 which 
is a cut-off that regularizes the loop integrals of the LS equation. 
Such a  
method proves rather successful 
since phase shifts and inelasticities are reproduced accurately up
 to 1200 MeV. 
The $f_0(980)$ and $a_0(980)$
resonances appear as poles of the $T$ matrix for I = 0 and 1, 
respectively, and their
widths and partial decay widths are very well reproduced.
In addition, one finds a pole when I = 0 at 
$\sqrt{s} \simeq 500$ MeV with a width of around 400 MeV, 
corresponding to the $\sigma$ meson, 
which was also found with similar properties with the IAM \cite{IAM}.

The appearance of the $f_0$ and $a_0$ is due to the
introduction of the $K \bar{K}$ channel in addition to $\pi \pi$ in
I = 0 and $\pi \eta$ in I = 1. These resonances disappear if the $K \bar{K}$
channel (not considered in \cite{truongdo,IAM}) is  omitted, while the 
$\sigma$ in
I = 0 is almost not affected. This explains why the $f_0$ and $a_0$
resonances did not show up in the IAM \cite{truongdo,IAM}.

The success of the scheme of ref.\cite{olset} in the scalar sector gives hopes
that it could be used in other channels. However, one soon realizes
that it does not reproduce properly the J = 1 sector. This looks
less surprising when one recalls that the $\Opc$ chiral Lagrangian 
can be reproduced with the resonance saturation hypothesis \cite{PiRa}.
That is, assuming that the actual values of the $\Opc$ parameters
are basically saturated by 
resonance exchanges between Goldstone bosons. In 
this way, one establishes a clear relation between the information contained
in the $\Opc$ Lagrangian and the resonances in the meson-meson sector,
particularly vector meson resonances, where the approach of \cite{PiRa} has
its stronghold. Indeed, the absence of the $\rho$ and $K^*$ 
in the
approach of \cite{olset}, which only uses the $\Opd$ Lagrangian, is an
indirect confirmation of the link between these resonances and the
$\Opc$ Lagrangian. 

The approaches of \cite{truongdo,IAM} and \cite{olset} seem
 complementary and one may wonder
whether there is a generalization of these methods, containing both them 
as limiting cases. An affirmative answer to this question was recently found 
and such a generalized method was proposed in \cite{prl}. The purpose of the present 
paper is to exploit the idea of \cite{prl} and obtain all the predictions of 
such approach in the meson-meson sector, like phase shifts, inelasticities, 
resonance properties, etc... At the same time we will establish the links 
between this scheme and $\chi PT$ at low energies. We also illustrate 
qualitatively, using a toy model, why the proposed method is so successful 
when dealing with amplitudes dominated by resonances.

\section{Unitary amplitude in coupled channels}

Let us write the partial wave decomposition of the meson-meson amplitude 
with definite isospin $I$ as

\begin{equation}
T_I = \Sigma_J (2 J + 1) \; T_{I J} \, P_J (cos \theta)
\label{TIJ}
\end{equation}
where $T_{I J}$ is the partial wave amplitude with isospin $I$  and angular 
momentum $J$. In each one of these channels there are several meson-meson 
states coupled to each other. In Table I, we have listed these states
for  the $J =  0, 1$ channels, which contain the most
relevant meson-meson information below 1 GeV. Note that  it is enough  
to take 
into account one or two states in each channel since we are neglecting
here, on the one hand,  multipion states which are only relevant for 
higher energies and, on
the other hand,  the $\eta\eta$ that appears for $(I,J)=(0,0)$. The 
influence of this state is 
rather small. We have checked it following the scheme of \cite{olset} and, 
although not zero, we found it small enough to omit it with the 
consequent simplicity in the general formalism. 

Hence, throughout the present work, $T_{IJ}$ will be either a $2 \times 2$
symmetric matrix when two states couple, or just a
number when there is only one state. In what follows we omit 
the $I$, $J$ labels
 and use a matrix formalism, which will be valid for the general case 
 of $n\times n$ matrices corresponding to $n$ coupled states.

The normalization of $T$ is such that
\begin{equation}
\frac{d \sigma}{d \Omega} = \frac{1}{64 \pi^2 s} \, \frac{k_f}{k_i}
\, \vert T_{if}\vert ^2
\end{equation}

\noindent
where $k_i$ and $k_f$ are, respectively,  the CM three momenta 
of the initial  and  final  state  and 
$s$ is the usual Mandelstam variable. Note that we have chosen a convention
for the sign of $T$ such 
that in an elastic amplitude $\Ima T \le 0$.

Unitarity in coupled channels implies

\begin{equation}
\Ima T_{if} = T_{in} \, \sigma_{nn} \, T^*_{nf}
\label{Tunit}
\end{equation}

\noindent
where $\sigma$ is a real diagonal matrix whose elements account 
for the phase space of the two meson intermediate states $n$ which are
physically accessible. With the normalization that 
we have chosen, $\sigma$ 
is given by the imaginary
part of the loop integral of two meson propagators in the $n$ state

$$
\sigma_{nn} (s) = \Ima G_{nn}(s)
$$

\begin{equation}
G_{nn}(s) = i \int \frac{d^4 q}{(2 \pi)^4} \;
\frac{1}{q^2 - m^2_{1 n} + i \epsilon} \;
\frac{1}{ (P - q)^2 - m^2_{2 n} + i \epsilon}
\label{G}
\end{equation}

$$
\Ima G_{nn}(s)=-\frac{k_n}{8 \pi \sqrt{s}} \theta (s-(m_{1n}+m_{2n})^2)
$$
\noindent
where $k_n$ is the on-shell CM momentum of the meson in the intermediate 
state $n$, P is the initial total four-momentum and
$m_{1 n}, m_{2 n}$ the masses of the two mesons in the state $n$.
An analytical expression for $G_{nn}(s)$ using a cut-off ($q_{max}$) 
regularization 
in the integral over $d^3q$ is shown in Appendix A. 

From eq.(\ref{Tunit}) we can  extract $\sigma$ and express it, 
in matrix form, as

\begin{eqnarray}
\Ima G &=& T^{- 1} \cdot \Ima T \cdot T^{* - 1}\nn \\
&=& \frac{1}{2 i} T^{- 1}\cdot (T - T^*)\cdot T^{* - 1} \nn\\
&=& \frac{1}{2 i} (T^{- 1 *} - T^{- 1}) = - \Ima T^{- 1}
\label{ImG}
\end{eqnarray}

Hence, 

\begin{eqnarray}
T^{- 1} &=& \Rea T^{- 1} - i \Ima G\nn \\
T &=& [\Rea T^{- 1} - i \, \Ima G]^{- 1}
\label{Tinverse}
\end{eqnarray}

This is a practical way to write the unitarity requirements of
eq.(\ref{Tunit}) 
which tells us that we only need to know $\Rea T^{- 1}$ 
since $\Ima T^{-1}$ is given by the phase space of the intermediate
physical states.

The next point is to realize that the $T$ 
matrix has poles associated to resonances, 
which implies that the standard perturbative 
evaluation of $\chi PT$ will necessarily fail
close to these poles. As a consequence, one might try to exploit the expansion 
of $T^{- 1}$, which will have zeros at the poles of T, and in 
principle does not present convergence problems. 
For illustrative purposes, we can use an analogy with the
function $\tan x$ when expanded around $x = 0$ ($x$ playing 
here the role of $p^2$ in the chiral expansion). This 
function has a pole at $x = \pi/2$. Its inverse, ${\cot} x$,  
has a Laurent 
expansion around $x = 0$ and a zero at $x = \pi/2$. 
However, inverting the expansion of 
${\cot}  x$ around $x=0$ for values of $x$ near $\pi/2$, provides 
a faster convergence than expanding directly $\tan x$ around that point. 
With this idea in mind let us expand $T^{-1}$ in powers 
of $p^2$ as one would do for $T$ using $\chi PT$:

$$
T \simeq T_2 + T_4 + ...
$$

\begin{equation}
T^{- 1} \simeq T_2^{- 1}\cdot [1 + T_4 \cdot T_2^{- 1} ...]^{- 1} 
\simeq T_2^{- 1}\cdot [1 - T_4 \cdot T_2^{- 1} ...]
\label{TIChPT}
\end{equation}

This expression requires the inversion $T_2$ which might not be 
invertible, as it happens, for instance in the (1, 1) channel. 
In order to avoid the use of  $T_2^{-1}$ we modify 
eq.(\ref{Tinverse}) by formally multiplying by $ T_2\cdot T_2^{- 1}$ on the
right and $T_2^{- 1}\cdot T_2$ on the left. All the steps are justified using 
the continuity of the functions involved in
the derivation, starting from a matrix close to $T_2$, which can be inverted.
Thus, eq.(\ref{Tinverse}) can be rewritten as

\begin{equation}
T = T_2\cdot [T_2 \cdot \Rea T^{- 1} \cdot T_2 - i T_2 \cdot \Ima G \cdot 
T_2]^{- 1}
 \cdot T_2
\label{preT}
\end{equation}

Now, using the expansion for $T^{-1}$ of eq.(\ref{TIChPT}) we find

\begin{equation}
T_2 \cdot \Rea T^{- 1} \cdot T_2 \simeq T_2 - \Rea T_4 + ...
\label{TReT}
\end{equation}
and recalling that

\begin{equation}
\Ima T_4 = T_2 \cdot \Ima G \cdot T_2
\end{equation}
we finally obtain, within the $\Opc$ approximation

\begin{equation}
T = T_2 \cdot [T_2 - T_4]^{- 1} \cdot T_2
\label{GenIAM}
\end{equation}

Note, as it is clear from eq.(\ref{preT}), that what we are expanding 
is actually 
$T_2 \cdot \Rea T^{-1} \cdot T_2$, which in our analogy would be equivalent to 
$x^2{\cot} \; x$, which is also convergent around $x=0$. 

In another context, the above equation can also be derived using Pad\'e
approximants \cite{Basde}.
This equation is a generalization to multiple coupled channels of the
IAM of ref.\cite{truongdo,IAM}. 
It makes the method more general and powerful and also allows 
to evaluate transition cross
sections as well as inelasticities. 

The coupled channel
result has additional virtues with respect to
the single channel IAM. Indeed, in this latter case the expansion of
eq.(\ref{TIChPT}) 
is meaningless if  $\vert T_2 \vert < \vert T_4 \vert$ or $T_2=0$ 
\cite{Pennington}. In particular, if $T_2$ vanishes,
 eq.(\ref{GenIAM}) yields $T = T^2_2 \cdot T_4^{- 1}$, which has a 
{\em double} zero, whereas the correct result would be
$T \simeq T_4$.  This indeed occurs in the $J=0$ partial waves
below threshold (Adler zeros). However, within the coupled 
channel formalism, if a matrix element, say $(T_{2})_{11}$, vanishes,
it is sufficient that $(T_{2})_{12} \neq 0$, since then eq.(\ref{GenIAM})
gives $(T)_{11} \simeq (T_{4})_{11}$, which is the
correct result. In conclusion, while the
single channel IAM gives a {\em double} zero whenever  $T_2=0$, the
coupled channel method  leads to {\em single} 
zeros close to the zeros of $T_2$. 

The single channel IAM has another related problem, 
since close to the Adler 
zero it presents an spurious pole when $T_{2}=T_4$. The coupled
channel method  also avoids this problem, although
it runs into a similar one when the determinant of the $T_2-
T_4$ matrix vanishes  below threshold. 
This happens indeed for $J=0$, $I=0$ around 
$\sqrt{s}\simeq 120$ MeV. Excluding the neighborhood of 
this zero of the determinant, we can still
recover from eq.(\ref{GenIAM}) the usual $\chi PT$ expansion,  
 $T \simeq T_2+T_4+...$ valid for low energies, typically $\vert 
\sqrt{s} \vert \, < \, 500$ MeV. In any case we concentrate here on 
results above the two pion threshold. 

It is now important to realize that 
eq.(\ref{GenIAM}) requires the complete evaluation of $T_4$, which
is rather involved when dealing with many channels, 
as it is the case here. 
Instead, we present a further approximation to eq.(\ref{GenIAM})
which turns out to be technically much simpler and rather accurate. 
In order
to illustrate the steps leading to our final formula, let us make 
before another approximation. 
Let us assume that through a suitable cut-off we can approximate

\begin{equation}
\Rea \, T_4 \simeq T_2 \cdot \Rea \, G \cdot T_2
\label{ReT4}
\end{equation}

In such a case we go back to eqs.(\ref{preT}, \ref{TReT}) and immediately write

\begin{equation}
T = [1 - T_2 \cdot G]^{- 1} \cdot T_2
\end{equation}
that is equivalent to 

\begin{equation}
T = T_2 + T_2 \cdot G \cdot T
\label{LS}
\end{equation}
which is a LS equation for the $T$ matrix, where $T_2$
plays the role of the potential. This is actually the approach followed in
ref. \cite{olset}.

There is a subtle difference between eq.(\ref{LS}) and the ordinary LS
integral equation. Indeed, eq.(\ref{LS}) is an algebraic equation
since $T_2$ and $T$ are factorized out of the integrals 
with their on-shell value. In contrast,
in the  ordinary LS 
equations, the $T_2 G T$ term is actually the integral of eq.(\ref{G}), 
including $T_2$ and $ T$ {\em inside} the integral, since both of them 
depend on $q$. Due to the structure of the $\Opd$ Lagrangian, it was shown
in  \cite{olset} that writing $T_2 (q)$ as $T_2 ^{on shell}(q)$ +
$T_2^{off shell}(q)$, the off-shell part renormalizes 
couplings and masses and hence it had to be omitted. 
Therefore $T_2$, and $T$ factorized {\em outside} the integral 
with their on-shell values.
As a consequence, the very same algebraic equation (\ref{LS}) was obtained.

As we have already commented, the approximation of eq.(\ref{ReT4}) 
leads to excellent results in the scalar channels.
However, as we mentioned in the introduction, the generalization to
$J\neq 0$ is not possible since basic information contained in the $\Opc$
chiral Lagrangian is missing in eq.(\ref{ReT4}). The obvious
solution is to add a term to eq.(\ref{ReT4}) such that
\begin{equation}
\Rea \, T_4 \simeq T_4^P + T_2 \cdot \Rea \, G \cdot T_2
\end{equation}
where  $T_4^P$ is the polynomial tree level contribution 
coming from the $\Opc$
Lagrangian, whose terms contain several free parameters, 
usually denoted $L_i$. 
Within our approach, these coefficients will be fitted to data
and denoted by $\hat{L}_i$ since they do not have to coincide with those used
in $\chi PT$, as we shall see. Actually, the $L_i$ coefficients depend on a  
regularization scale ($\mu$). In our scheme this scale 
dependence appears through
the cut-off. 

In addition, there are also differences between our
renormalization scheme and that of standard $\chi PT$. Indeed, 
our approach considers the iteration of loop
diagrams in the s-channel, but neglects loops in the u or t channels. 
However,
the smooth structure of these terms for the physical s-channel, since 
we are far away from the associated singularities, allows
them to be approximately reabsorbed when fitting the $\hat{L}_i$
coefficients. Concerning tadpoles, they would be
exactly reabsorbed in the $\hat{L}_i$ in the equal mass case. 
Therefore, when masses are different, we are omitting terms
proportional to differences between the actual masses squared and
an average mass squared. Thus all these contributions 
will make the $\hat L_i$ differ from the $L_i$, although we expect
them to be of the same order.

This way of dealing with tadpoles has an additional advantage.
Apart from the usual tadpole diagrams that would also appear in
standard $\chi PT$ there are some additional tadpole terms. They come from the 
determinant of the $SU(3)$ metric that should be included in the 
path integral measure in order to make the generating
functional $SU(3)$ covariant \cite{covariant}.
With dimensional regularization such contributions vanish, but 
that is not the case when using a cutoff regularization \cite{Tararu}. 
Nevertheless, we have just described how tadpoles are absorbed within 
our approximation and thus we do not have to calculate them.

With these approximations our calculations have been considerably 
simplified at the expense of losing some precision at low energies 
 with respect to the full $\Opc$ $\chi PT$ calculation. 
As far as we are mostly interested in resonance behavior as well
as higher energies this is not very relevant.
Nevertheless, if the complete $\Opc$ 
calculations were available, we could directly use eq.(\ref{GenIAM}), and
have both an accurate low energy description and a good coupled
channel unitarity behavior.

Using eqs.(\ref{preT}) and  (\ref{TReT}),  our 
final formula for the $T$ matrix is given by

\begin{equation}
T = T_2 \cdot [T_2 - T_4^P - T_2\cdot G \cdot T_2]^{- 1} \cdot T_2
\label{ourT}
\end{equation}

\section{Toy model}

In order to illustrate how the method works, we take a simple
case of one channel and one amplitude around a resonance which we assume to
know exactly. That is,

\begin{equation}
T = \frac{a p^2}{q^2 - M^2 + i \, 2 M \Gamma}
\label{toy}
\end{equation}
where $p^2$ is an invariant quantity, of dimension momentum squared, 
related to 
the momenta or masses of the pseudoscalar mesons, q the total four-momentum 
of the meson pair and $2 M \Gamma = - a p^2 \Ima G$. 
The above equation satisfies unitarity
exactly as can be seen by using eq.(\ref{ImG}).

To ${\cal O}(k^2)$, $k\equiv p,\; q$, we have

\begin{equation}
T_2 = - a \frac{p^2}{M^2}
\end{equation}
whereas at ${\cal O}(k^4)$ we have

\begin{equation}
\Rea T_4 = - \frac{a p^2 q^2}{M^4} \equiv T_2 \frac{q^2}{M^2}
\end{equation}

Then, using eq.(\ref{GenIAM}) we find

\begin{eqnarray}
T & =& \frac{T_2^2}{T_2 - \Rea T_4 - i T_2 \Ima G T_2} =
- \frac{a p^2}{M^2 (1 - \frac{q^2}{M^2} + i a \frac{p^2}{M^2} \Ima G)} \nn \\
&=& \frac{a p^2}{q^2 - M^2- i a p^2 \Ima G}
\end{eqnarray}

So, as we can see, in this particular case the IAM
leads to the exact result for the $T$ matrix, eq.(\ref{toy}). The result is 
exact
here because $T_2 \cdot \Rea T^{- 1}\cdot  T_2$ is an ${\cal O}(k^4)$ function  and hence
the expansion up to ${\cal O}(k^4)$ in eq.(\ref{TReT}) is exact. However,
the structure of eq.(\ref{toy}) is that of a meson propagator of an unstable
particle like the $f_0, a_0, \rho, K^*$, etc... resonances. This could 
justify 
why the scheme which we propose works
even better that one could naively anticipate, at least for resonant 
channels. 

The above argumentation uses the same power counting in momenta as 
$\chi PT$, 
but presumes that the ${\cal O}(k^2)$ amplitude comes from the exchange of a 
resonance. This seems to be in conflict with \cite{PiRa}, where it is shown 
that resonance exchange contribution shows up  at higher orders. However, 
when 
taking into account requirements of short distance behavior of QCD, these 
two points can be reconciled. In fact, this has been shown, in \cite{
guerrer}, where a classical vector meson dominance expression for the pion 
form factor is obtained, in the same lines as eq.(\ref{toy}), starting 
from chiral 
Lagrangians and imposing those QCD constraints at short distances and 
the large $N_c$ limit.

In relation to the previous arguments, 
 the link between unitarized $\chi PT$ and vector meson 
dominance has also been discussed in \cite{VMD}.

\section{The matrix elements of $T_2$ and $T_4$.}

The lowest order chiral Lagrangian is given by

\begin{equation}
{\cal L}_2 = \frac{f^2}{4} \langle \partial_{\mu} U^{\dagger} \partial^{\mu} U +
M (U + U^{\dagger})\rangle
\end{equation}
where $f$ is the pion decay coupling and  $\langle \rangle$ stands for the trace of the
$3 \times 3$ matrices build out of $U (\Phi)$ and $M$.

\begin{equation}
U (\Phi) = \exp (i \sqrt{2} \Phi / f)
\end{equation}
where $\Phi$ can be expressed in terms of the Goldstone boson fields as

\begin{equation}
\Phi (x) \equiv
\left(
\begin{array}{ccc}
\frac{1}{\sqrt{2}} \pi^0 + \frac{1}{\sqrt{6}} \eta & \pi^+ & K^+ \\
\pi^- & - \frac{1}{\sqrt{2}} \pi^0 + \frac{1}{\sqrt{6}} \eta & K^0 \\
K^- & \bar{K}^0 &  - \frac{2}{\sqrt{6}} \eta
\end{array}
\right)
\end{equation}

The mass  matrix $M$ is given by

\begin{equation}
M = 
\left(
\begin{array}{ccc}
m^2_{\pi} & 0 & 0 \\
0 & m^2_{\pi} & 0 \\
0 & 0 & 2 m^2_K - m^2_{\pi}
\end{array}
\right)
\end{equation}
where we have assumed the isospin limit $m_u = m_d$.

The $\Opc$ Lagrangian is given by

\begin{eqnarray}
{\cal L}_4 &=& L_1 \langle \partial_{\mu} U^{\dagger} 
\partial^{\mu} U \rangle^2 +
L_2 \langle \partial_{\mu} U^{\dagger} \partial_{\nu} 
U \rangle \langle\partial^{\mu}
U^{\dagger} \partial^{\nu} U \rangle \nn\\
&+& L_3 \langle \partial_{\mu} U^{\dagger} \partial^{\mu} 
U \partial_{\nu} U^{\dagger}
\partial^{\nu} U \rangle + 
L_4 \langle \partial_{\mu} U^{\dagger} \partial^{\mu} U \rangle
\langle U^{\dagger} M + M^{\dagger} U \rangle \nn\\
&+& L_5 \langle \partial_{\mu} U^{\dagger} \partial^{\mu} 
U (U^+ M + M^+ U)\rangle
+ L_6 \langle U^{\dagger} M + M^+ U\rangle^2 \nn\\
&+& L_7 \langle U^{\dagger} M - M^{\dagger} U \rangle^2 +
L_8 \langle M^{\dagger} U M^{\dagger} U + U^{\dagger} M U^{\dagger} M \rangle
\end{eqnarray}
where the terms which couple to external sources are omitted \cite{GaLe}. 

The states with definite isospin, with the phases $ \vert \pi^+ \rangle = 
- \vert1,1 \rangle $, $
\vert K^- \rangle = - \vert 1/2 - 1/2\rangle$, are given by

\begin{eqnarray*}
&&{\bf I = 0}, \nn\\
&&\vert K \bar{K} \rangle= - \frac{1}{\sqrt{2}} \vert K^+ (\vec{q}) 
K^- (- \vec{q}) +
K^0 (\vec{q}) \bar{K} \, ^0 (- \vec{q}) \rangle \nn\\
&&\vert\pi \pi\rangle = - \frac{1}{\sqrt{6}} \vert \pi^+ (\vec{q}) 
\pi^- (- \vec{q})
+ \pi^- (\vec{q}) \pi^+ (- \vec{q}) + \pi^0 (\vec{q}) 
\pi^0 (- \vec{q}) \rangle \nn\\[2ex]
&&{\bf I = 1, I_3 = 0}, \nn\\
&&\vert K \bar{K} \rangle = - \frac{1}{\sqrt{2}} \vert K^+ (\vec{q}) 
K^- (- \vec{q}) +
K^0 (\vec{q}) \bar{K} \, ^0 (- \vec{q}) \rangle \nn\\
&&\vert\pi \eta \rangle = \vert\pi^0 (\vec{q}) 
\eta (- \vec{q}) \rangle \nn\\
&&\vert\pi \pi \rangle = - \frac{1}{2} \vert\pi^+ (\vec{q}) \pi^- 
(- \vec{q}) -
\pi^- (\vec{q}) \pi^+ (- \vec{q}) \rangle \nn\\[2ex]
&&{\bf I= 2, I_3 = 2}, \nn\\
&&\vert\pi \pi\rangle = \frac{1}{\sqrt{2}} \vert\pi^+ (\vec{q}) 
\pi^+ (- \vec{q}) \rangle \nn\\[2ex]
&&{\bf I = 1/2, I_3 = 1/2}, \nn\\
&&\vert K \pi \rangle = - \vert \frac{\sqrt{2}}{3} \pi^+ (\vec{q}) 
K^0 (- \vec{q}) +
\frac{1}{\sqrt{3}} \pi^0 (\vec{q}) K^+ (- \vec{q}) \rangle \nn\\
&&\vert K \eta \rangle = \vert K^+ (\vec{q}) \eta (- \vec{q}) 
\rangle \nn\\[2ex]
&&I = 3/2, I_3 = 3/2, \nn\\
&&\vert K \pi \rangle = - \vert K^+ (\vec{q}) \pi^+ (- \vec{q}) \rangle
\end{eqnarray*}

We should note that in the states of identical particles 
we have included an extra
$1/\sqrt{2}$ factor
in the normalization. This is done to ensure that the resolution of the
identity  gives unity (recall that 
$\Sigma_q \vert\pi^0 (\vec{q}) \pi^0 (- \vec{q}) \rangle \langle \pi^0 (\vec{q}) 
\pi^0 (- \vec{q})\vert=2$ with the states $\pi^0 (\vec{q}) \pi^0 (- \vec{q})$
normalized to unity). This normalization yields the ordinary
unitarity formulae, eq.(\ref{Tunit}), which we are using to  extract phase shifts and
inelasticities. However, we should return to the proper normalization
 at the end in order to obtain the physical amplitudes.

The amplitudes which we obtain are  compiled in Appendix B. The projection
over each partial wave J is done by means of

\begin{equation}
T_{IJ} = \frac{1}{2} \int^1_{- 1} \, P_J (\cos \theta) \, T_I (\theta) d( \cos \theta) 
\end{equation}

In the case of two coupled channels, $T_{IJ}$ is a $2 \times 2$ matrix
whose elements, $(T_{IJ})_{ij}$ are related to 
$S$ matrix elements through the equations (omitting 
the $I,J$ labels)

\begin{equation}
\begin{array}{l}
(T)_{11} = - \frac{8 \pi \sqrt{s}}{2 i p_1} \, [(S)_{11} - 1] \; , \;
(T)_{22} = - \frac{8 \pi \sqrt{s}}{2 i p_2} \, [(S)_{22} - 1] \\
(T)_{12} = (T)_{21} = - \frac{8 \pi \sqrt{s}}
{2 i \sqrt{p_1 p_2}} \, (S)_{12}
\end{array}
\label{TandS}
\end{equation}
with $p_1$, $p_2$ the CM momenta of the mesons in state $1$ or $2$
respectively. The $S$ matrix has the structure \cite{WeIs}

\begin{equation}
S =
\left[
\begin{array}{cc}
\eta e^{2 i \delta_1} & i (1 - \eta^2)^{1/2} \, 
e^{i (\delta_1 + \delta_2)} \\
i (1 - \eta^2)^{1/2} \, e^{i (\delta_1 + \delta_2)} & \eta e^{2 i \delta_2}
\end{array}
\right]
\end{equation}

\noindent
where $\delta_1$ and $ \delta_2$ are the phase shifts for the 
elastic $1\rightarrow 1$ and $2\rightarrow 2$ 
processes (for instance, $\bar{K} K\rightarrow \bar{K} K$ and 
$\pi \pi\rightarrow \pi\pi$ 
in $(I,J)=(0,0)$) and $\eta$ is the
inelasticity.

It is interesting to note that, by means of $(T)_{11}$ 
and $(T)_{22}$ one can 
determine $\eta, \delta_1$ and $ \delta_2$, and hence 
the $(T)_{12} = (T)_{21}$ matrix elements 
are redundant. We determine them from our coupled 
equations and verify that the
structure of eq.(\ref{TandS}) is satisfied, which is another check of
the coupled channel unitary that we have imposed from the beginning.

\section{Results}

We have carried out a fit to the data,  which is shown in figs. 1 to
7, using as free parameters the  $\hat L_{i}$ with  $i=1,2,3,4,5,7$ and 
$2 \hat L_6 + \hat L_8$.  
The cut-off is fixed to $q_{max}=1.02$ GeV. The values which we obtain are 
shown in Table II.  By comparing them with the standard values for 
the $L_i$ coefficients 
obtained in $\chi PT$ at the scale $\mu=2\, q_{max}/\sqrt{e}$ (see appendix 
A.2) we see that 
they are of the same order.

We show first the results on phase shifts 
and inelasticities in the different channels 
and later on we discuss about the pole positions, widths and partial decay 
widths. 

\subsection{Phase shifts and inelasticities}

We will now go in detail through the results in each (I,J) channel.

\subsubsection{Channel (0,0)}

As we can see 
in eq.(\ref{TandS}) we have three independent magnitudes 
$\delta_{1}, \delta_{2}$ and 
$\eta$. In figs.(1.a) and (1.c) we show the $\delta_1$ and  
$\delta_2$ corresponding to 
$\pi\pi\rightarrow\pi\pi$ and $K\bar{K}\rightarrow K\bar{K}$ 
elastic scattering. In fig.(1.b) we plot the phase shift for 
$K \bar{K}\rightarrow\pi\pi$. 
This is actually $\delta_1+\delta_2$, which is therefore  redundant
information.  
However, there are data for this process but not for elastic $K \bar{K}$, 
and that is why we are plotting   $\delta_1+\delta_2$. 
The agreement with experiment is good, with small discrepancies in the 
$K\bar{K}\rightarrow \pi\pi$ phase shifts. In 
fig.(1.a) we see a bump around 600 MeV which is due to
the $\sigma$ 
resonance, whose associated pole appears around 
$442 -i 225$ MeV, as we shall see 
below. The fast raise in the phase shift at 1 GeV is caused by
the $f_0$ pole around $980 - i 14$ MeV, which translates in
an apparent mass of $\simeq 980$ MeV and a 30 MeV width. Small 
discrepancies with data start showing up around 
1.2 GeV. The omission of the $\eta\eta$ and four meson states
should limit the validity of the 
approach at high energies since then these 
channels start being relevant. 

\subsubsection{Channel (1,1)}

In fig.(2.a) we display the 
$\pi\pi\rightarrow\pi\pi$ 
phase shifts which clearly show the $\rho$ meson. 
The perfect coincidence of the results with the very precise 
data indicate that both the position and the width of the $\rho$
are very well described. In fig.(2.c) 
we show the phase shifts for $K \bar{K}\rightarrow K\bar{K}$ scattering, for 
which there 
are no data. As we can see, they are very small, which implies
a weak $K\bar{K}$ interaction. Therefore the $\delta_1+\delta_2$ phase 
shift of $K \bar{K}\rightarrow\pi\pi$ 
is essentially that of  
$\pi\pi\rightarrow\pi\pi$. The fact that the inelasticity is practically 
one, 
indicates that there is almost no mixture of $\pi\pi$ and $K\bar{K}$. 
This feature makes the $\rho$ to behave as a pure $\pi\pi$ 
elastic resonance. 
That is why the single channel IAM gave essentially the same results 
as obtained here \cite{truongdo}.

\subsubsection{Channel (2,0)}

The $I=2$ $\pi\pi$ scattering contains only one state as shown in Table I. 
In fig.(3) we show the resulting phase shifts, whose agreement with
experimental data is remarkably good up to 1.2 GeV

\subsubsection{Channel (1,0)}
 
In fig.(4.a) the $\pi\eta\rightarrow\pi\eta$ phase 
shifts are shown. Those of 
$K\bar{K}\rightarrow K\bar{K}$ are plotted in fig.(4.b) and the
inelasticities in fig.(4.c). 
In the latter, it can be seen that there is an appreciable mixture 
between $\pi\eta$ 
and $K\bar{K}$ above $K\bar{K}$ threshold. In fig.(4.d) we compare a mass 
distribution for $\pi\eta$ around the region of the $a_0$ resonance. 
The data are 
obtained from \cite{Amsterdam} using the $K^-p\rightarrow 
\Sigma^+(1385)\pi\eta$ 
reaction, whose cross section (following  \cite{Flatte}) can be written as

\begin{equation}
\frac{d\sigma}{dm}=C {\vert t \vert}^2 q
\label{Flatte}
\end{equation}
where $m$ is the $\pi^-\eta$ invariant mass, $q$ the $\pi$ momentum in
the $\pi^- \eta$ CM 
frame, $t$ the $\pi^-\eta \rightarrow \pi^- \eta$ scattering 
amplitude and $C$ a normalization 
constant. We observe a fairly good agreement with the experimental numbers.

\subsubsection{Channel (1/2,0)}

The two coupled states are now $K\pi$ and  $K\eta$. In Fig.(5.a) we
plot the  phase shifts for $K \pi \rightarrow K \pi$. The agreement 
of the results
with the data is quite good, although a bit on 
the upper part. The results and the data show a broad bump, which 
is related to the presence of a pole which appears around 
$770- i 250$ MeV. Such a resonance, whose existence has been claimed
in a recent data analysis \cite{japon},
is predicted in quark models of $q^2 \bar{q}^2$ systems \cite{Jaffe} 
and is usually denoted 
by $\kappa(900)$. 
This resonance bears some similarity with the $\sigma$ in the (0,0)
$\pi\pi$ elastic scattering channel, which is also very
broad. Finally, the $K\eta
\rightarrow K\eta$ phase shifts are small as shown in fig.(5.c) and 
the inelasticities given in fig.(5.d) are not distant from unity.
This fact indicates  
a small mixture of $K \pi$ with $K \eta$.

\subsubsection{Channel (1/2,1)}

In this case we also find a 
resonance in fig.(6.a), analogous to the $\rho$, but in the $K \pi$
system.  This resonant state, known as the $K^*(892)$,
 is as clean as the $\rho$, and the agreement of our results with the 
data is remarkably good over the whole range of energies up to 1.2
GeV. In fig.(6.c) we plot the $K \eta \rightarrow K \eta$ phase shifts,
  which 
are very small. Finally, in fig.(6.d) we can notice that $\eta \approx
1$ which means that there is practically no mixture of 
$K \pi$ and $K \eta$ in this channel. This justifies the success of 
\cite{truongdo} reproducing this resonance using only the $K \pi$ state and 
elastic unitarity.

\subsubsection{Channel (3/2,0)}

In fig.(7) we show the $K\pi$ phase shifts. As we can see in the 
figure, the agreement with the data is quite good up to about $1.2$ GeV.

\vskip .5cm

The channel (3/2,1) in $K\pi$ (see table I) is such that 
$T_2=0$, since there is only S-wave there. In this case our method cannot
be applied, as discussed above, and we should just take the $T_4$
contribution. 
That also happens for the $J=2$ 
 channels, since the structure of $T_2$, which is $\Opd$, 
 is a linear combination of $s$, 
$t$, $u$ and squared masses. Therefore there is only $J=0, 1$ 
in $T_2$, but not 
$J=2$. Hence, the lowest contribution 
can only be obtained from the $T_4$ terms 
and our method 
has nothing to improve there with respect to $\chi PT$. The phase shifts in 
these channels 
are small and have been discussed in \cite{truongdo}. Hence we omit
any further discussion, 
simply mentioning that the agreement with data found in \cite{truongdo} 
is fairly good. 

There is another interesting result in the (0,1), channel which is the 
appearance of a pole around 990 MeV, that we show in fig.(8). 
Below 1.2 GeV there are two 
resonances with such quantum numbers. They are the
$\omega$ and the $\phi$, which fit well within the $q\bar{q}$ 
scheme, with practically ideal mixing, 
as 
$\frac{1}{\sqrt{2}}(u\bar{u}+d\bar{d})$ and $s\bar{s}$, respectively. Hence, 
the $\omega$ would almost decouple from $K\bar{K}$ and then we
should not expect it to appear in our scheme with only the $K \bar{K}$ 
channel. The three pion channel, into which the $\omega$ mostly decays, is not 
considered in our approach, restricted to two meson states. In contrast, the 
$\phi$ couples 
strongly to the $K \bar{K}$ system.  It seems then natural to identify
the above mentioned pole with the $\phi$. Indeed, it is only 
30 MeV below its real mass, 
1020 MeV (which means a relative deviation of only 3 $\%$). 
Due to this shift 
towards lower energy this resonance appears below the
$K\bar{K}$  threshold and this is why we find no width at all. 
Nevertheless, as far as its
physical width is only $\simeq 4$ MeV, it seems plausible that
that the small coupling to three pions (an  
OZI suppressed coupling of third class) which we are not taking 
into account, 
could be enough to improve the agreement between the position of our $\phi$ 
resonance and its real mass and width.  

\subsection{Pole positions, widths and partial decay widths.}

We will now look for the poles of the $T$ matrix in the complex
plane, that should appear in the unphysical Riemann sheets ( 
the conventions taken are those of \cite{olset}, which can be easily 
induced from the analytical expressions of Appendix A). 
Let us remember that the mass and the width of a Breit-Wigner
resonance are related to the position of its
complex pole by $\sqrt{s_{pole}}\simeq M - i \Gamma/2$, but this
formula does not hold for other kind of resonances.
 In table II we give the results for the pole positions as well as the
apparent or ``effective''
masses and widths that can be estimated from phase shifts and mass
distributions in scattering processes. Note that such ``effective'' masses 
and widths depend on the physical process.

We shall make differentiation between the $\rho$ and $K^*$, which are clean 
elastic Breit-Wigner resonances, and the rest. For the $\rho$ and $K^*$ their mass is  
given by the energy at which $\delta=90^0$ and the width is taken from the 
 phase shifts slope around $\delta=90^0$, by means of 

\begin{equation}
\Gamma_R=\frac{M_R^2-s}{M_R} \tan \delta (s)
\label{pela}
\end{equation} 

We also saw that, in practice,  the $\rho$ and $K^*$ 
only couple to $\pi\pi$ and $K\pi$,
respectively. The $\sigma$ decays only to $\pi\pi$ 
and the $\kappa$ only to $K\pi$ due to 
phase space and dynamical suppression of other channels (see fig.(5d)). The 
case of the $f_0$ and $a_0$  is
different, since they can decay either to $\pi\pi$ or 
$K \bar{K}$ (the $f_0$),
and $\pi\eta$ or  $K\bar{K}$ (the $a_0$). In order to determine the partial 
decay widths of these resonances we follow the procedure 
of \cite{olset}, where we show that, assuming a Breit Wigner shape for the 
amplitudes around the resonance pole, the partial decay widths are given by

\begin{eqnarray}
\Gamma_{R,1} &=& \frac{1}{16 \pi^2} \int_{E_{min}} ^{E_{max}} 
dE \frac{q}{E^2} 
4 M_R \Ima T_{11}   \nn\\
\Gamma_{R,2} &=&  -\frac{1}{16\pi^2} 
\int_{E_{min}} ^{E_{max}} dE \frac{q}{E^2} 
4 M_R \frac{(\Ima T_{21})^2}{\Ima T_{11}}  
\label{oset}
\end{eqnarray}
where $E$ stands for the total CM energy 
of the meson-meson system, $q$ is the 
momentum of one meson in the CM and 
the labels $1,2$ stand for $K\bar{K}$, $\pi\pi$ in the case of the 
$f_0$ and 
$K \bar{K}$, $\pi\eta$ in the case of the $a_0$. The masses of the final 
mesons are $m_1$, $m_2$. The upper 
limit in the integral, $E_{max}$, 
is $\simeq M_R+\Gamma_R$ where $\Gamma_R$ is the total width \cite{olset} 
and $E_{min}=M_R-\Gamma_R$, unless the threshold energy ($m_1+m_2$) for the 
decay is bigger than that quantity, in which case $E_{min}=m_1+m_2$. 
In this way we largely avoid 
the contribution of the backgrounds in the amplitudes. 
One caveat must be raised concerning eq.(\ref{oset}), which 
was already pointed out in the study of the 
$f_0\rightarrow\gamma \gamma$ decay
\cite{os}. The subtlety is that around this resonance the phase shifts 
(see fig.(1.a)) are of the order of $90^0$, due 
to the background coming from the 
broad $\sigma$ pole. This background 
makes the $f_0\rightarrow\pi\pi$ coupling constant to appear effectively 
multiplied by a $\pi/2$ phase ($i$ factor) and in this way the $T_{12}$ 
amplitude around the $f_0$ looks like an ordinary Breit Wigner multiplied 
by $i$. This means that the real 
part has a peak around the resonance and the imaginary part changes sign. 
In this case the arguments used in \cite{olset} and \cite{os} 
lead to a trivial 
modification in  $\Gamma_{R,2}$,
where $\Ima T_{12}$ should be substituted by $\Rea T_{12}$.

It is also very  instructing to see the representation of the poles in 
a three dimensional plot. In fig.(9) 
we are showing on the left the imaginary part of the (0,0)
$\pi\pi\rightarrow\pi\pi$ scattering amplitude on the second Riemann
sheet. It is possible to see very clearly the appearance of two poles
that correspond to the $\sigma$ and the $f_0$ resonances. The former
is located at $442 - i 227 $ and thus is very far away from the real
axis, which implies a huge effective width. In contrast, the other pole
is located at $994 -i  14 $ MeV accordingly to the narrow width of
the $f_0$ resonance. 

Apart from the position of the poles, there is an additional piece of
information which also determines the observed
shape of a resonance. It also explains some of the differences between
the ``effective'' masses and the real part of the pole position.
On the right of fig.(9) we give a contour
plot, 
again of the imaginary part of the (0,0) amplitude in the second Riemann
sheet. Notice that both poles are oriented differently, almost
transversally, on the complex plane. On the one hand, the $f_0$ pole is 
oriented almost 
perpendicularly to the real axis, which is the relevant one in this work. 
As a consequence, in the positive real axis,
the imaginary part of the amplitude
first grows rapidly and then drops very fast again, giving rise to the
dramatic variation of the phase shift typical of resonances. 
A similar orientation is found for the $\rho, K^*$ and $ a_0$
resonances too.
On the other hand, the $\sigma$ pole is oriented so that in the real
axis we only see a slow and smooth increase, but almost no decrease,
of the imaginary part. That is also the case of the $\kappa$
resonance. This feature, together with the fact that both the $\sigma$
and the $\kappa$ are very far from the real axis explains why it is so
hard to establish firmly their existence and their physical
parameters.

Finally, in fig.(10) we present a very detailed contour plot of the
$\rho$ and $a_0$  poles. Both of them are almost perpendicular to the
real axis, but the former is tilted clockwise, whereas the
latter is tilted anti-clockwise. Let us now remember that 
 the real part of the pole position, roughly, should give
us the apparent mass of the resonance. However,  the lines of maximum
gradient of each pole cross the real axis at a point which is slightly 
different
from the real part of its position. Therefore, those poles 
rotated clockwise, as
the $\rho$ or the $K^*$, have an apparent mass a little bit  higher than
that given by the pole position. In contrast
those tilted anti-clockwise, yield a resonance whose mass is
somewhat lower that the one obtained from the pole. That is the case
of the $f_0$ and the $a_0$.

\section{Conclusions and Outlook}

We have used a coupled channel unitary approach, 
together with the dynamical 
information  contained in the $\Opd$ and $\Opc$ 
chiral Lagrangian, which allows us to study the 
meson-meson interaction 
up to about 1.2 GeV. 
This non-perturbative method generates poles in the complex 
plane corresponding to physical resonances. 
We have used the experimental 
information available to make a fit of the 
$\Opc$ Lagrangian coefficients. 
These are $\hat L_i$, $i=1,2,3,4,5,7$ 
and  $2 \hat L_6+\hat L_8$, whose actual values depend
on the cut-off that we have used to regularize 
divergent one loop integrals.
With those seven degrees 
of freedom we are able to fit, up to 1.2 GeV,
 all the experimental information
in seven meson-meson channels. Each one of this channels
consists of two phase shifts and an  inelasticity.
Moreover, in our results, we obtain the 
position and widths, partial decay widths, etc....
of all the resonances that appear in those channels below 1.2 GeV. 
Apart from the standard $f_0$, 
$a_0$, $\rho$, $K^*$ resonances, 
we find poles in the $T$ matrix 
for the $\sigma$ in the $\pi\pi$ $I=J=0$ 
channel and for $\kappa$ in the (1/2,0) channel, 
both them very broad. 

The method has proved very efficient  
to extend the ideas of chiral symmetry at energies 
beyond the realm of applicability of $\chi PT$. 
However, at energies higher than 1.2 GeV,
the limitations of the model show up, since,
among other things, we have  restricted ourselves
 only to two meson states.
The restrictions in the space of states precluded the appearance of 
the $\omega$ resonance which couples dominantly
to three pions. However, the $\phi$ resonance which 
couples strongly to $K\bar{K}$ does appear
in the scheme, although slightly shifted  towards lower energies. 
Presumably, by including the $\phi$
coupling to three pions, although very small, 
it should be enough to 
shift the mass to its correct place.

One of the formal weakness of the approach is that loops in crossed
channels, as well as some tadpole contributions, are
not explicitly included in the calculation. In practice, their
effect can be reabsorbed in the fit of the $\Opc$ parameters, 
whose values can then be different from those 
obtained for the standard low energy $\chi PT$ approach. 

This approximation could be improved by using eq.(\ref{GenIAM}) 
with the full  $\Opc$ $\chi PT$ calculation, which 
includes one loop 
in crossed channels and the tadpoles. 
This would allow a more 
straightforward comparison with standard $\chi PT$
as well as a better accuracy in the low energy results. 
Although such calculations are welcome and there is indeed some
work in progress 
\cite{PacoJA}, they are far more involved to calculate and use.

Applications of the method to other physical problems are also 
in order. Indeed, it  can be easily 
extended to deal with processes where meson pairs appear in the 
initial or final state, like meson pair photoproduction \cite{os}. 
It looks likely that it could also prove useful
describing the meson-nucleon interaction \cite{Angels} complemented
with Heavy Baryon Chiral Perturbation Theory. 
In addition,
the method, non perturbative in nature, is equally well
suited to study the meson-meson interaction 
in a nuclear medium where there has been some speculation about the 
appearance of bound $\pi\pi$ pairs \cite{Schuck}.

Finally it seems that the approach could be extended to the
effective chiral Lagrangian  description of
the Standard Model Strongly Interacting 
Symmetry Breaking Sector, where the single channel approach 
has already been applied \cite{LHCres}.

\section*{Acknowledgments}

We are grateful to A. Dobado for discussions concerning the 
tadpole contributions and for his careful reading of the manuscript.
Two of us, J.A.O and E.O. would like  to thank the kind hospitality
of the Complutense University of Madrid. J.R.P. wishes to 
thank the hospitality of the University of Valencia and the
SLAC Theory Group. This work
was partially supported by DGICYT under contracts PB96-0753 and
AEN93-0776. J. A. O.  and J.R.P. acknowledge financial support from the
Generalitat Valenciana and the Ministerio de Educaci\'on y Cultura, 
respectively.

\appendix

\section{Analytical formula for $G(s)$: Relation between cut-off 
and  dimensional renormalization}

In this appendix we are showing the relationship between our
regularization scheme and dimensional regularization, which is the
usual one when dealing with $\chi PT$. 

\subsection{Analytical formula for $G(s)$ with a cut-off regularization}

In the general case with different masses, $M_1$ and $M_2$

\begin{eqnarray}
G(s) & = & \frac{1}{32 \pi^2}\Bigg[-\frac{\Delta}{s}\log\frac{M_1^2}{M_2^2}
+\frac{\nu}{s}
\left\{ \log\frac{s-\Delta+\nu\sqrt{1+\frac{M_1^2}{q_{max}^2}}}
{-s+\Delta+\nu \sqrt{1+\frac{M_1^2}{q_{max}^2}}} +
\log\frac{s+\Delta+\nu\sqrt{1+\frac{M_2^2}{q_{max}^2}}}
{-s-\Delta+\nu \sqrt{1+\frac{M_2^2}{q_{max}^2}}} \right\} \nonumber \\   
& & +2\frac{\Delta}{s}\log\frac{1+\sqrt{1+\frac{M_2^2}{q_{max}^2}}}{
1+\sqrt{1+\frac{M_2^2}{q_{max}^2}}} -2\log \Bigg[ \left(1+\sqrt{1+
\frac{M_1^2}{q_{max}^2}}\right)\left(1+\sqrt{1+\frac{M_2^2}{q_{max}^2}}\right) 
\Bigg]
\nonumber \\
& & +\log\frac{M_1^2 \, M_2^2}{q_{max}^4} \Bigg]
\label{ana1}
\end{eqnarray}  
where $\nu=\sqrt{(s-(M_1+M_2)^2)(s-(M_1-M_2)^2)}$ and $\Delta=M_1^2-
M_2^2$. In the case of equal 
masses, $M_1=M_2=m$, the above formula reduces to

\begin{equation}
G(s)=\frac{1}{(4 \pi)^2} \left[ \sigma \log \frac{\sigma \sqrt{1+
\frac{m^2}{q_{max}^2}}+1}{\sigma \sqrt{1+\frac{m^2}{q_{max}^2}}-1}
-2 \log\left\{ \frac{q_{max}}{m}\left(1+\sqrt{1+\frac{m^2}{q_{max}^2}} 
\right) \right\}\right]
\label{ana2}
\end{equation}
where now, $\sigma=\sqrt{1-4m^2/s}$.

The numerical evaluation 
of the principal part of eq.(\ref{G}) is also performed as an additional check.

\subsection{Relation between the cut-off and the dimensional
  regularization scale}

In order to obtain the relationship between the cut-off and 
the renormalization 
scale $\mu$ let us consider, for the sake of simplicity, 
the case with equal masses (the same result is obtained with different masses 
but the formulas 
are more cumbersome). As far as we are going to compare the same function 
calculated in different ways, let us denote by $G^C(s)$ the $G(s)$ calculated 
with 
a cut-off regularization and $G^D(s)$ the one calculated with dimensional 
regularization. In this latter case we have

\begin{equation}
G^D(s)=\frac{1}{(4\pi)^2} \left[ \frac{1}{\hat\epsilon}-2+\log \, m^2
+\sigma \log\frac{\sigma+1}{\sigma-1} \right]
\label{Dim}
\end{equation}
where $1/\hat \epsilon=1/\epsilon-\log(4\pi)+\gamma$ with $D=4+2\epsilon$.

The scale $\mu$ in $G^D(s)$ appears through the inclusion of the $L_i$ 
\cite{ChPT} at $\Opc$

\begin{equation}
L_i=L_i^r(\mu)+\Gamma_i \lambda
\end{equation}

where $L_i^r(\mu)$ is the renormalized value of $L_i$ at the $\mu$ scale, 
$\Gamma_i$ is just a number and 

\begin{equation}
\lambda=\frac{1}{32 \pi^2}\left[ \frac{1}{\hat{\epsilon}}+\log \mu^2-1 \right]
\end{equation}

The $\log \mu^2$, and its companion $\frac{1}{\hat{\epsilon}}-1$, are 
incorporated in $G^D(s)$ so that at the end one has a logarithm of the dimensionless 
quantity $m^2/\mu^2$. In this way  we rewrite $G^D(s)$ as:

\begin{equation}
G^D(s)=\frac{1}{(4\pi)^2} \left[ -1+\log  \frac{m^2}{\mu^2}
+\sigma \log\frac{\sigma+1}{\sigma-1} \right]
\label{GD1}
\end{equation}

We expand eq.(\ref{ana2}) in powers of $m^2/q_{max}^2$ to compare with the 
cut-off regularization, as follows

\begin{eqnarray}
G^{C}(s) &=&\frac{1}{(4\pi)^2}\left[ -2 \log\frac{2 q_{max}}{m}+\sigma \log
\frac{\sigma +1}{\sigma-1} +{\cal O} \left( \frac{m^2}{q_{max}^2} \right) 
\right]
\nonumber \\
&=&\frac{1}{(4\pi)^2}\left[-1+\log\, e +\log\frac{m^2}{4 \, q_{max}^2}
+\sigma \log \frac{\sigma+1}{\sigma-1}+ {\cal O} \left(\frac{m^2}{q_{max}^2}\right)\right]
\nonumber \\
&=&\frac{1}{(4\pi)^2}\left[-1+\log \frac{m^2 \, e}{4q_{max}^2}+
\sigma \log \frac{\sigma+1}{\sigma-1} + {\cal O}
\left(\frac{m^2}{q_{max}^2}\right)\right]
\label{gc1}
\end{eqnarray}

 Then comparing eqs.(\ref{GD1}) and (\ref{gc1}) one has:

\begin{equation}
\label{relationship}
\mu=\frac{2\,q_{max}}{\sqrt{e}}\simeq 1.2 \, q_{max}
\end{equation}

Hence, to our cut-off $q_{max}\simeq1 $ GeV would 
correspond a $\mu =\,1.2$  
GeV dimensional regularization scale.
In Table II, we have listed the values of the $\hat L_i$ 
parameters and those of standard $\chi PT$ scaled to $\mu =\,1.2$ GeV. 
As it is explained in the text,
in our fit we have neglected the crossed channel diagrams and 
we have treated tadpoles differently. The effect of these contributions
is effectively reabsorbed in our $\hat L_i$ parameters, hence some differences 
between the $\hat{L}_i$ and $L_i$ parameters should be expected and this is 
indeed the case as can be seen in Table II. Note that, even if we had used 
the complete $\Opc$ $\chi PT$ calculations,
these parameters could be different, since they have been obtained from a fit 
over a much wider range of energies than it is used in $\chi PT$ and higher 
order contributions have been included.

Finally, note that the terms 
${\cal O}( m^2/q_{max}^2)$ in eq.(\ref{gc1}) 
yield ${\cal O} (p^6)$, or higher, contributions 
and that is why they are not included in 
$G^D(s)$.

It is also worth stressing that the relationship 
of eq.(\ref{relationship}) 
is independent of the physical process and channel 
since the function $G(s)$ 
appears in all them in the same way. 
  
\section{Amplitudes}

We have used the following formulae in our calculations. Note that,
as it has been explained in the text,
we have an overall sign of difference with the definitions in \cite{GaLe},
as well as a $1/2$ factor for those amplitudes with identical particles.

\subsubsection{Masses and decay constants}

\begin{eqnarray}
f_{\pi} &=& f_0 \left[ 1+\frac{4 m_{\pi}^2}{f_0^2}(L_5+L_4)+\frac{8m_K^2}{f_0
^2}L_4 
\right] \nn \\
f_K&=&f_0 \left[ 1+\frac{4m_K^2}{f_0^2}(L_5+2L_4)+\frac{4m_{\pi}^2}{f_0^2}
L_4 \right] 
\nn \\
f_{\eta}&=&f_0 \left[ 1+\frac{4 m_{\eta}^2}{f_0^2}L_5+\frac
{8m_K^2+4m_{\pi}^2}{
f_0^2}
L_4 \right]
\end{eqnarray}

\vskip .5 cm

\begin{eqnarray}
m_{\pi}^2&=&m_{0\;\pi}^2 \left[ 1 +\frac{8 m_{\pi}^2}{f_0^2}(2L_6+
L_8-L_4-L_5)+\frac{16m_K^2}{f_0^2}(2L_6-L_4) \right]  \nn \\
m_K^2&=&m_{0\;K}^2 \left[ 1 +\frac{16 m_K^2}{f_0^2}(2L_6+
L_8-L_4-\frac{1}{2}L_5)+\frac{8m_{\pi}^2}{f_0^2}(2L_6-L_4) \right]
\end{eqnarray}
where the $0$ subscript refers to bare quantities.
\vskip .5cm

\subsubsection{$\pi \pi\rightarrow \pi \pi$ scattering}

The definite isospin amplitudes $T^{(I)}$ are obtained 
from just one amplitude $T$:
\begin{eqnarray}
T^{(0)}(s,t,u)&=& (3T(s,t,u)+ T(t,s,u) + T(u,t,s))/2\nn\\
T^{(1)}(s,t,u)&=& (T(t,s,u) - T(u,t,s))/2\nn\\
T^{(2)}(s,t,u)&=& (T(t,s,u) + T(u,t,s))/2\nn\\
\end{eqnarray}
where $T=T_2+T_4$ is given by:
\begin{eqnarray}
T_2&=& \frac{\mpi-s}{\fpi} \\
T_4&=&-\frac{4}{\fpc}\left\{
(2L_1+L_3)(s-2\mpi)^2+L_2\left[(t-2\mpi)^2+(u-2\mpi)^2\right]\right.\nn\\
&&\spa\left.+2(2L_4+L_5)\mpi(s-2\mpi)+4(2L_6+L_8)m_\pi^4\right\} \nn
\end{eqnarray}
which have been obtained at tree level from $\La_2$ and $\La_4$, respectively.

\subsubsection{$K \pi \rightarrow K \pi $ scattering}

Using crossing symmetry, we can write the $I=1/2$ amplitude in
terms of that with $I=3/2$, as
\begin{equation}
T^{(1/2)}(s,t,u)=\frac{3}{2}T^{(3/2)}(u,t,s)-\frac{1}{2}T^{(3/2)}(s,t,u)
\end{equation}
where
\begin{eqnarray}
T_2^{(3/2)}&=& \frac{s-(\sq)}{2f_\pi f_K} \\
T_4^{(3/2)}&=&-\frac{2}{\fpi\fk}\left\{ (4L_1+L_3)(t-2\mpi)(t-2\mk)
+2L_2(\sq-s)^2\right.\nn\\
&+&\left.(2L_2+L_3)(\sq-u)^2+4L_4\left[(\sq)t-4\mpi\mk\right]\right.\nn\\
&+&\left.L_5\left[(\sq)(\sq-s)-4\mpi\mk\right]+8\mpi\mk(2L_6+L_8)\right\}\nn
\end{eqnarray}
once more, they have been obtained, respectively, 
from $\La_2$ and $\La_4$ at tree level.

\vskip .5cm

\subsubsection{$K \oK\rightarrow K \oK$ scattering}

The definite isospin amplitudes can be written just in terms of two:
\begin{eqnarray}
T^{(0)}(s,t,u)&=&T^{+-+-}(s,t,u)+T^{\bar{0}0+-}(s,t,u)\\
T^{(1)}(s,t,u)&=&T^{+-+-}(s,t,u)-T^{\bar{0}0+-}(s,t,u)\nn
\end{eqnarray}
where $T^{+-+-}$ is the amplitude for $\Kma\Kme\rightarrow\Kma\Kme$, whose
respective $\Opd$ and $\Opc$ contributions are
\begin{eqnarray}
T_2^{+-+-}(s,t,u)&=&\frac{u-2\mk}{\fk}\\
T_4^{,+-+-}(s,t,u)&=&-\frac{4}{\fkc}
\left\{2L_2(u-2\mk)^2+(2L_1+L_2+L_3)
\left[(s-2\mk)^2+(t-2\mk)^2\right]\right.\nn\\
&&\spa\left. -2u\,\mk(2L_4+L_5)+8m_K^4(2L_6+L_8) \right\}\nn
\end{eqnarray}
whereas $T^{\bar{0}0+-}$ is the amplitude for $\oK^0K^0\rightarrow\Kma\Kme$,
which is given by
\begin{eqnarray}
T_2^{\bar{0}0+-}(s,t,u)&=&\frac{u-2\mk}{2\fk}\\
T_4^{\bar{0}0+-}(s,t,u)&=&-\frac{2}{\fkc}
\left\{(4L_1+L_3)(s-2\mk)^2+(2L_2+L_3)(t-2\mk)^2+2L_2(u-2\mk)^2\right.\nn\\
&&\left.+2s\,\mk(4L_4+L_5)+\frac{2}{3}t\, \mk(2L_4+L_5)
-8m_K^4(2L_4+L_5-2L_6-L_8)\right\}\nn
\end{eqnarray}

\subsubsection{$\pi \pi\rightarrow K \oK$ scattering}

Again, we can use crossing symmetry to obtain, from $K \pi \rightarrow K \pi $,
the definite isospin amplitudes $T_I$ of this process:
\begin{eqnarray}
T^{(0)}&=&\frac{\sqrt{3}}{2}\left(T^{(3/2)}(u,s,t)+T^{(3/2)}(t,s,u)\right)\\
T^{(1)}&=&\frac{1}{\sqrt{2}}\left(T^{(3/2)}(u,s,t)-T^{(3/2)}(t,s,u)\right)\nn
\end{eqnarray}

\subsubsection{$K \eta \rightarrow K \eta $ scattering}

This process is pure $I=1/2$. 
We obtain the following contributions to the amplitude:
\begin{eqnarray}
T_2(s,t,u)&=&\frac{6\me+2\mpi-9t}{12f_\eta f_K}\\
T_4(s,t,u)&=&-\frac{1}{3\fk\fe}\left\{
2(t-2\mk)(t-2\me)(12L_1+5L_3)+\left[
(u-\me-\mk)^2\right.\right.\\
&+&\left.\left
(s-\me-\mk)^2\right](12L_2+L_3)
+2(t-2\mk)\left[11\mk(2L_4+L_5)-\mpi(L_4+3L_5)\right]\right.
\nn\\ &+&
4m_K^4\left[3(2L_4+L_5)+32(L_6+L_7+L_8)\right]
+2(t-2\me)\left[9\mk(2L_4+L_5)+\mpi(3L_4-L_5)\right]
\nn\\&+&
4m_\pi^4(16L_7+8L_8-L_5)+6L_5m_\eta^4
-t\left[\mk(24L_4+7L_5)+2\mpi(6L_4-L_5)+9L_5\me\right]
\nn\\&+&
6\me\mk(4L_4+L_5)+2\me\mpi(6L_4+L_5)
+2\left.
\mk\mpi\left[6L_4+L_5-8(2L_6+7L_8+12L_7)\right]\right\}\nn
\end{eqnarray}

\subsubsection{$K \eta \rightarrow K \pi $ scattering}

The $I=1/2$ amplitude can be obtained as follows:
\begin{equation}
T^{(1/2)}(s,t,u)=\sqrt{\frac{3}{2}}T_{\oK^0 \eta\rightarrow\Kme\pi^+}(s,t,u)
\end{equation}                                                    
The $\Opd$ and $\Opc$ contributions to $\oK^0 \eta\rightarrow\Kme\pi^+$ are
\begin{eqnarray}
T_2(s,t,u)&=&\frac{\sqrt{6}\left[8\mk+3\me+\mpi-9t\right]}{36f_K f_\eta}\\
T_4(s,t,u)&=&-\frac{\sqrt{2/3}}{3\fk\fe}\left\{
3L_3\left[ 2(t-\mpi-\me)(t-2\mk)
-(s-\mk-\mpi)(s-\mk-\me)\nn\right.\right.\\
&-&\left.(u-\mk-\mpi)(u-\mk-\me)\right]
+ L_5\left[(t+\mpi-\me)(7\mk-5\mpi)\right.\nn\\
&+&\left.4m_K^2(3t-3\mpi-\me)
+2(t-2\mk)(\mk+\mpi)+4(m_\pi^4-m_K^4)\right]\nn\\
&+&\left.+16(2L_7+L_8)(m_\pi^4-2m_K^4+\mk\mpi)\right\}\nn
\end{eqnarray}
\subsubsection{$\eta\pi\rightarrow\eta\pi$}

This channel is pure $I=1$ isospin. The amplitude is given by
\begin{eqnarray}
T_2(s,t,u)&=&\frac{-\mpi}{3f_\eta f_\pi}\\
T_4(s,t,u)&=&-\frac{4}{3\fk\fe}\left\{ (t-2\mpi)(t-2\me)(6L_1+L_3)
+4t\, L_4(\mpi +2\mk) \right.\nn\\
&+&(3L_2+L_3)\left[(s-\mpi-\me)^2+(u-\mpi-\me)^2\right]+
m_\pi^4(4L_4-L_5-8L_6+32L_7+12L_8)\nn\\
&-&\left.16\mk\mpi(L_4-2L_6+2L_7)-3\mpi\me(4L_4+L_5)\right\}
\end{eqnarray}


%
%

\newpage
\begin{table}[h]
\begin{center}
{\bf{\small{Table I: Physical states used in the different $I,J$ channels}}}

\vskip .5cm

\begin{tabular}{|c|c|c|c|c|c|}
\hline
& I=0 & I=1/2 & I=1 & I=3/2 & I=2 \\
\hline       
J=0 & $\begin{array}{c} \pi\pi \\ K \bar{K}\\ \end{array} $&
$\begin{array}{c} K \pi \\ K \eta \\ \end{array}$ & $\begin{array}{c} 
\pi \eta \\ K \bar{K} \\ \end{array}$ & $K \pi$ & $\pi\pi$ \\
\hline 
J=1 & $K \bar{K}$ & $\begin{array}{c} K \pi \\ K \eta \\ \end{array}$ & $
\begin{array}{c} \pi \pi \\ K \bar{K} \\ \end{array}$ & & \\
\hline
\end{tabular}
\end{center}
\end{table}

\begin{table}[h]
\begin{center}
{\bf{\small{Table II: Fit parameters $\hat L_i \cdot 10^{3}$ and comparison 
with the $L_i^r \cdot 10^{3}$ of $\chi PT$ }}}

\vskip .5cm

\begin{tabular}{|c|c|c|c|c|c|c|c|}
\hline
$q_{max}=1.02$  
GeV 
& $ \begin{array}{c} \\ \hat L_1 \\ \\ \end{array} $ 
& $ \begin{array}{c} \\ \hat L_2 \\ \\ \end{array} $ 
& $ \begin{array}{c} \\ \hat L_3 \\ \\ \end{array} $ 
& $ \begin{array}{c} \\ \hat L_4 \\ \\ \end{array} $ 
& $ \begin{array}{c} \\ \hat L_5 \\ \\ \end{array} $ 
& $ \begin{array}{c} \\ 2 \hat L_6 + \hat L_8 \\ \\ \end{array} $
& $ \begin{array}{c} \\ \hat L_7 \\ \\ \end{array} $ \\ 
\hline            
& $ \begin{array}{c} \\ 0.5  \\ \\ \end{array} $ 
& $ \begin{array}{c} \\ 1.0  \\ \\ \end{array} $ 
& $ \begin{array}{c} \\ -3.2 \\ \\ \end{array} $ 
& $ \begin{array}{c} \\ -0.6 \\ \\ \end{array} $ 
& $ \begin{array}{c} \\ 1.7  \\ \\ \end{array} $ 
& $ \begin{array}{c} \\ 0.8  \\ \\ \end{array} $
& $ \begin{array}{c} \\ 0.2  \\ \\ \end{array} $ \\
\hline
$\mu=1.2$
GeV 
& $ \begin{array}{c} \\ L_1^r            \\ \\ \end{array} $ 
& $ \begin{array}{c} \\ L_2^r            \\ \\ \end{array} $ 
& $ \begin{array}{c} \\ L_3             \\ \\ \end{array} $ 
& $ \begin{array}{c} \\ L_4^r            \\ \\ \end{array} $ 
& $ \begin{array}{c} \\ L_5^r             \\ \\ \end{array} $ 
& $ \begin{array}{c} \\ 2L_6^r + L_8^r      \\ \\ \end{array} $
& $ \begin{array}{c} \\ L_7             \\ \\ \end{array} $\\
\hline               
& $ \begin{array}{c} \\ 0.1  \\ \pm 0.3 \\ \\ \end{array} $ 
& $ \begin{array}{c} \\ 0.9  \\ \pm 0.3 \\ \\ \end{array} $ 
& $ \begin{array}{c} \\ -3.5 \\ \pm 1.1 \\ \\ \end{array} $ 
& $ \begin{array}{c} \\ -0.7 \\ \pm 0.5 \\ \\ \end{array} $ 
& $ \begin{array}{c} \\ 0.4 \\ \pm 0.5 \\ \\ \end{array} $ 
& $ \begin{array}{c} \\ 0.0  \\ \pm 0.3 \\ \\ \end{array} $
& $ \begin{array}{c} \\ -0.4 \\ \pm 0.2 \\ \\ \end{array} $ \\   
\hline
\end{tabular}
\end{center}
\end{table}

\begin{table}[h]
\begin{center}
{\bf{\small{Table III. Masses and partial widths in MeV}}}
\vskip .5cm
\begin{tabular}{|c|c|c|c|c|c|c|}
\hline
  $ \begin{array}{c} {\rm Channel}  \\ (I,J)       \\ \end{array} $ 
&  Resonance 
& $ \begin{array}{c}  {\rm Mass }   \\ {\rm from \, pole}  \\ \end{array} $
& $ \begin{array}{c}  {\rm Width}   \\ {\rm from \, pole}  \\ \end{array} $
& $ \begin{array}{c}  {\rm Mass}   \\ {\rm effective} \\ \end{array} $
& $ \begin{array}{c}  {\rm Width}  \\ {\rm effective} \\ \end{array} $ 
& $ \begin{array}{c}  {\rm Partial}\\ {\rm Widths}\\ \end{array} $ \\ 
\hline  
$(0,0)$
& $\sigma $
& $442$
& $454$
& $\approx$ $600$ 
& $very$ $large$
&$\pi\pi - 100\%$\\
\hline
$(0,0)$
& $f_0(980)$
& $994$
& $28$
& $\approx$ $980$ 
& $\approx$ $30$
& $\begin{array}{c} \pi\pi -65\%\\ K \bar{K} - 35\% \\ \end{array}$ 
\\ 
\hline
$(0,1)$
& $\phi(1020)$
& $980$
& $0$
& $980$
& $0$
& 
\\
\hline
$(1/2,0)$
& $\kappa$
& $770$
& $500$
& $\approx$ $850$
& $very$ $large$
& $K \pi - 100\%$
\\
\hline
$(1/2,1)$
& $K^*(890)$
& $892$
& $42$
& $895$
& $42$
& $K \pi -  100\% $
\\
\hline
$(1,0)$
& $a_0(980)$
& $1055$
& $42$
& $980$
& $40$
& $\begin{array}{c}\pi\eta - 50 \% \\ K \bar{K} -  50\% \\ \end{array}$
\\ 
\hline
$(1,1)$
& $\rho(770)$
& $759$
& $141$
& $771$
& $147$
& $\pi\pi - 100\%$
\\ 
\hline
\end{tabular}
\end{center}
\end{table}

\newpage

\begin{figure}
\hbox{
\psfig{file=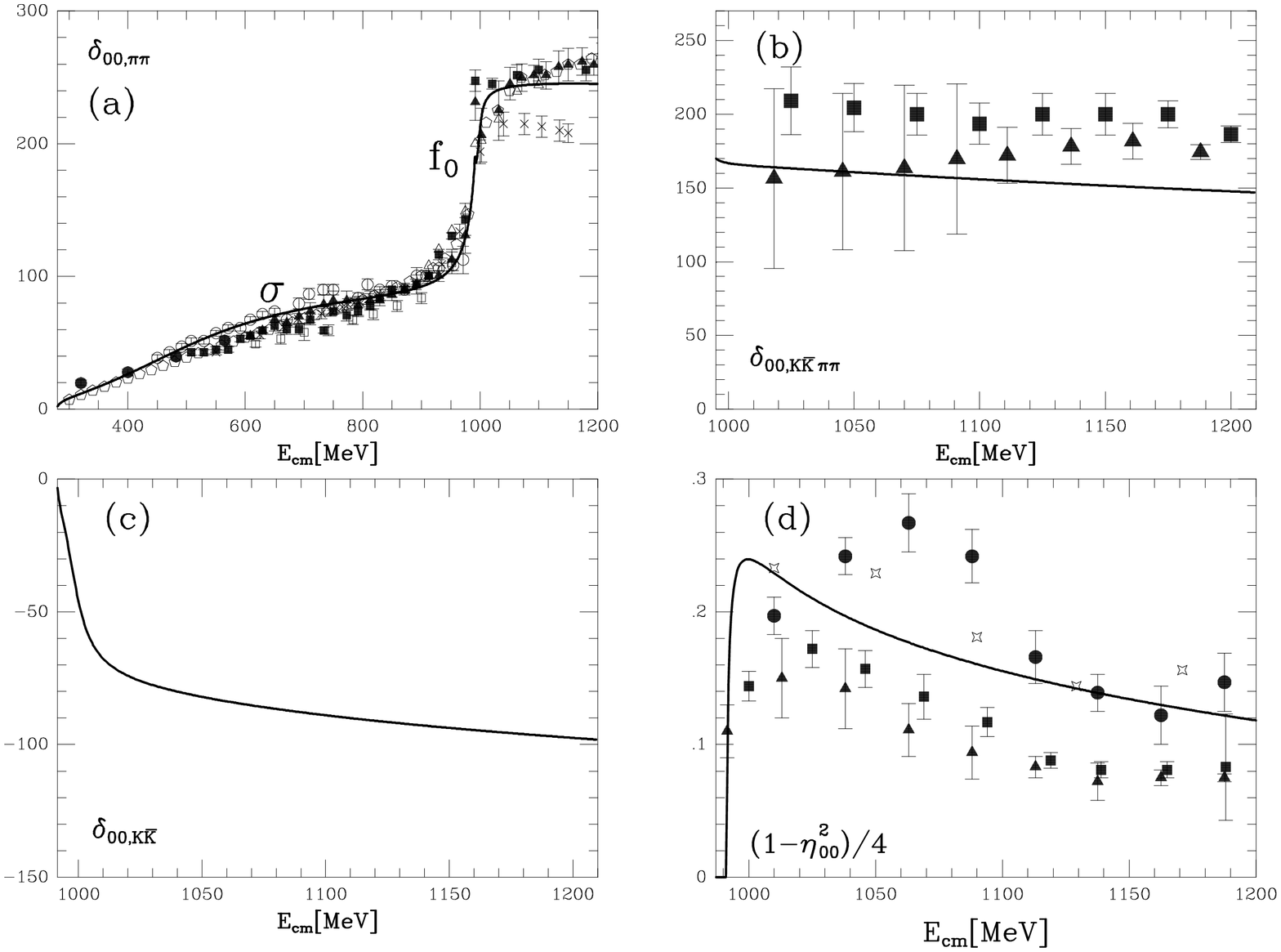,width=14cm,angle=-90}}
  {\footnotesize Fig.1: Results in the $I=J=0$ channel. (a) phase shifts for
  $\pi\pi\rightarrow\pi\pi$ as a fraction of the CM energy of the
  meson pair: full triangle \cite{Hyam}, open circle \cite{Esta2}, 
  full square \cite{Gay2}, open triangle \cite{Gay3}, open square \cite{Gay4} 
  (all these are analysis of the same experiment \cite{Grayer}), 
  cross \cite{Proto}, full circle \cite{Man}, empty pentagon \cite{Fro}. 
  (b) phase shifts for $K\bar{K}\rightarrow \pi\pi$: full square \cite{Cohen}
  , full triangle \cite{Martin}. (c) Phase shifts for $K \bar{K} \rightarrow 
  K \bar{K}$. (d)
  Inelasticity: results and data for $(1-\eta^2)/4$: starred square 
  \cite{Fro}, full square \cite{Cohen}
  , full triangle \cite{Martin}, full circle \cite{Linden}.}
\end{figure}

\begin{figure}
\hbox{
\psfig{file=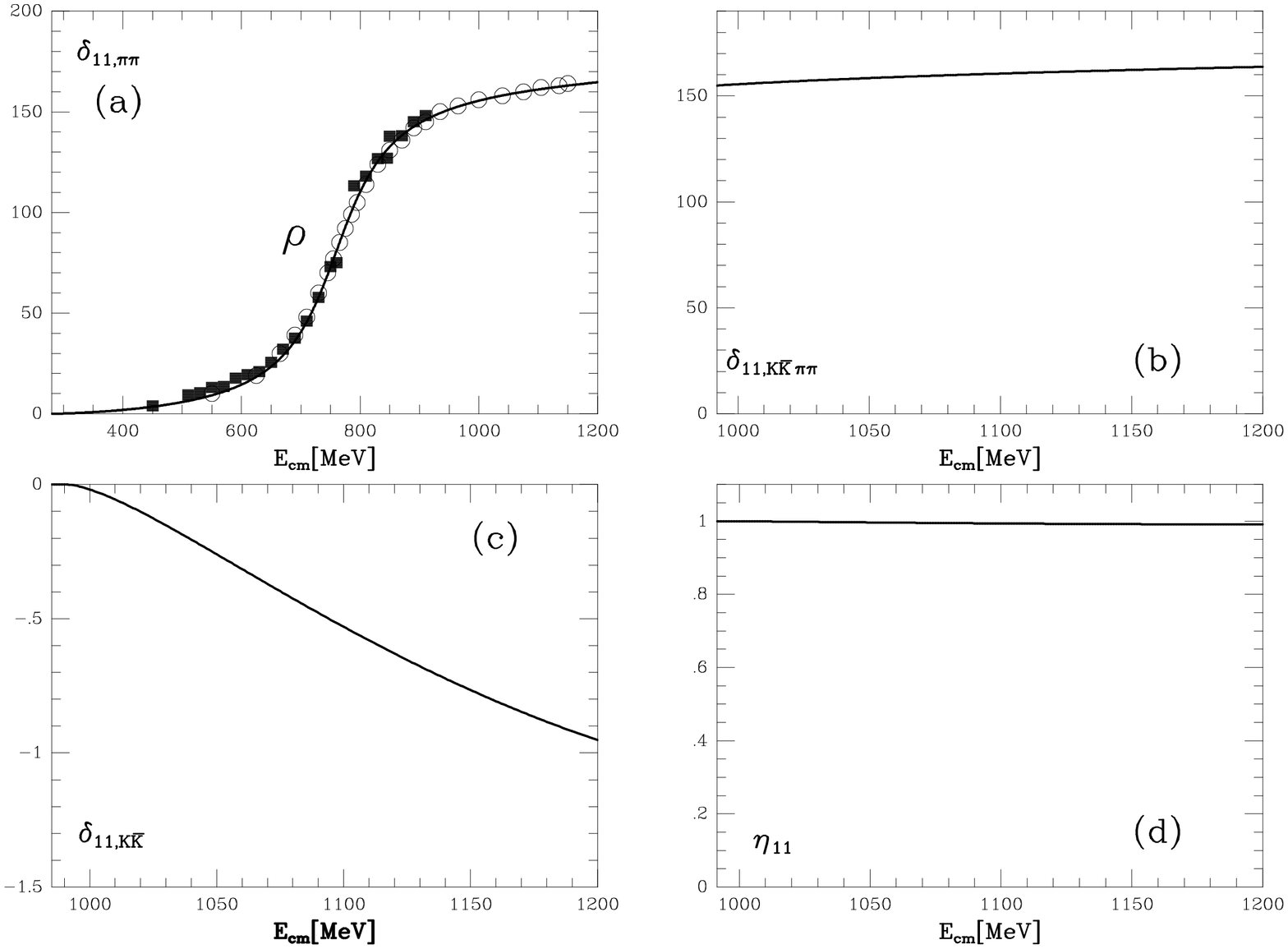,width=14cm,angle=-90}}
  {\footnotesize  Fig. 2: Results in the $I=J=1$ channel. (a) phase shifts for
  $\pi\pi\rightarrow\pi\pi$. Data: open circle \cite{Proto}, black square
  \cite{Esta}. (b), (c) same as in fig. 1. (d) inelasticity.}
\end{figure}

\begin{figure}
    \hbox{ \psfig{file=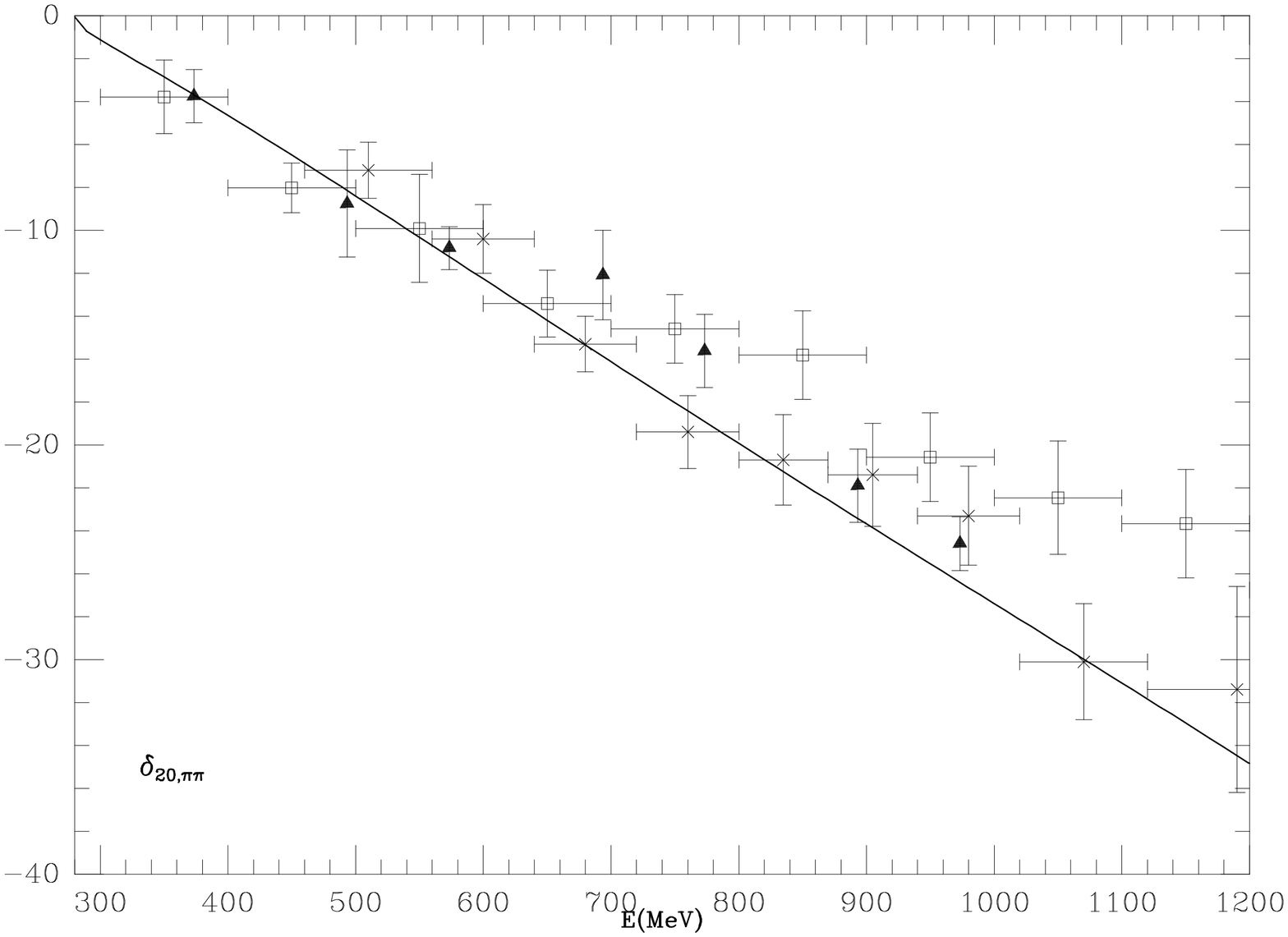,width=8cm,angle=-90}}
  {\footnotesize Fig. 3: Phase shifts for $\pi\pi\rightarrow\pi\pi$ in the 
  $I=2$, $J=0$ channel. Data: cross \cite{Rosselet}, empty square \cite{
  Sche}, full triangle \cite{Jan}.}
\end{figure}

\begin{figure}
\hbox{
\psfig{file=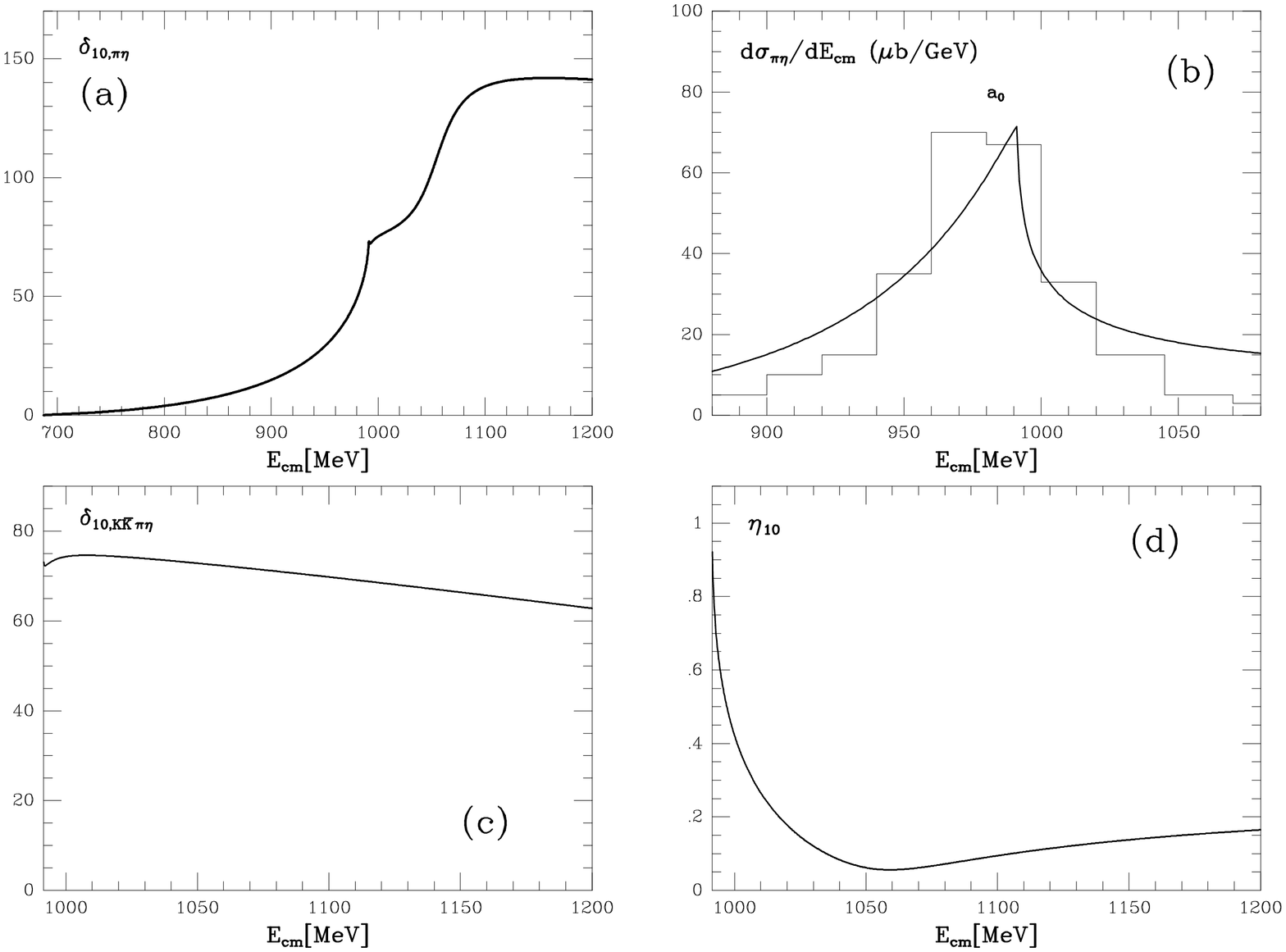,width=14cm,angle=-90}}
  {\footnotesize Fig. 4: Results in the $I=1$, $J=0$ channel. (a) phase shifts 
  for $\pi\eta\rightarrow \pi\eta$. (b) Invariant mass distribution for
  $\pi\eta$ data from \cite{Amsterdam}. (c) Phase shifts for $K \bar{K}
  \rightarrow \pi \eta$. (d) inelasticity.}
\end{figure}

\begin{figure}
\hbox{
\psfig{file=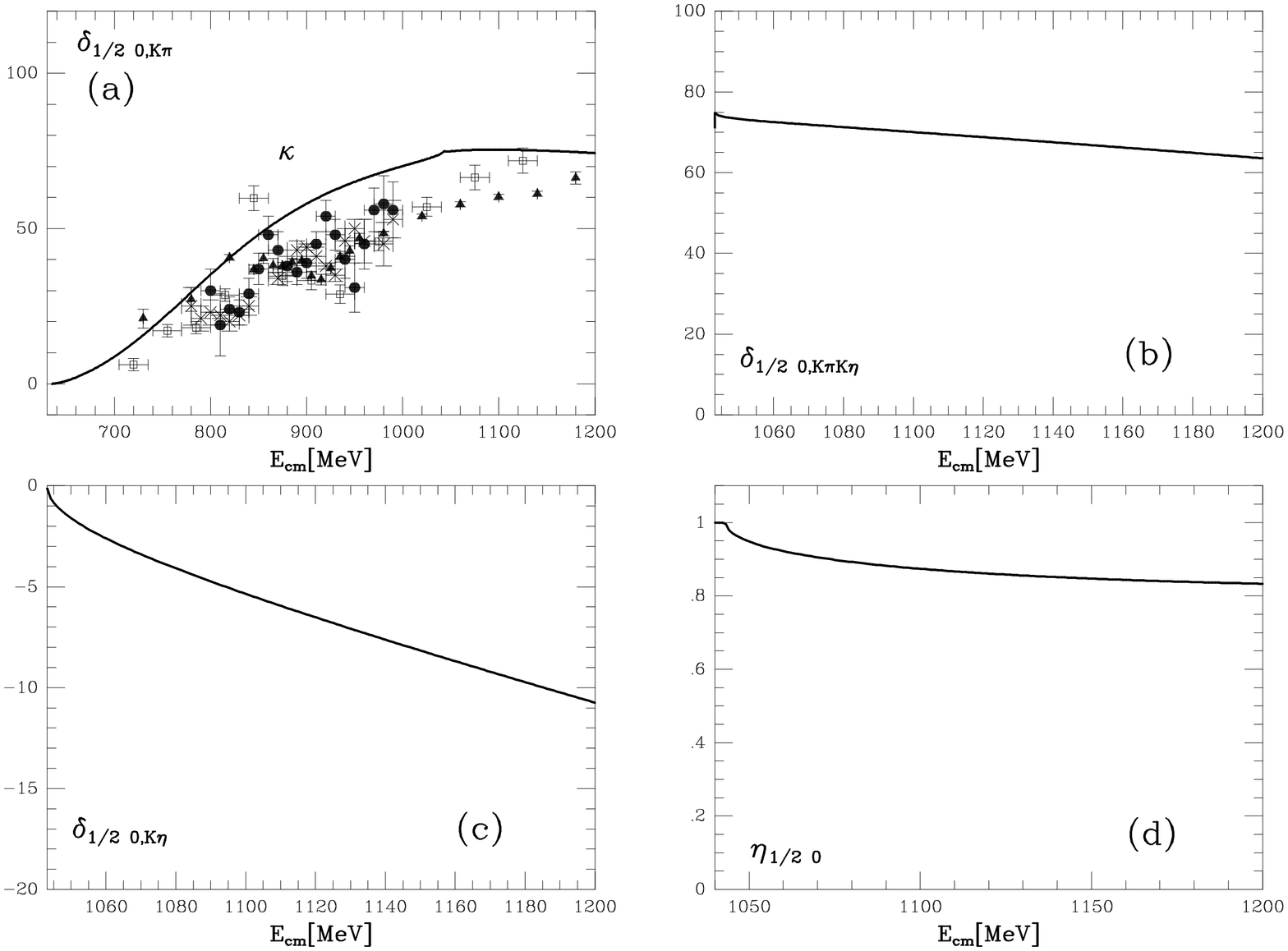,width=14cm,angle=-90}}
  {\footnotesize Fig. 5: Results in the $I=1/2$, $J=0$ channel. (a) phase shifts for
  $K\pi\rightarrow K\pi$. Data: full circle \cite{MeAn71}, cross \cite{BiDu72}
  , open square \cite{BaBa75}, full triangle \cite{EsCa78}. (b) phase shifts 
  for $K\pi\rightarrow K\eta$. (c) phase shifts for $K\eta \rightarrow
  K\eta$. (d) inelasticity.}
\end{figure}

\begin{figure}
\hbox{
\psfig{file=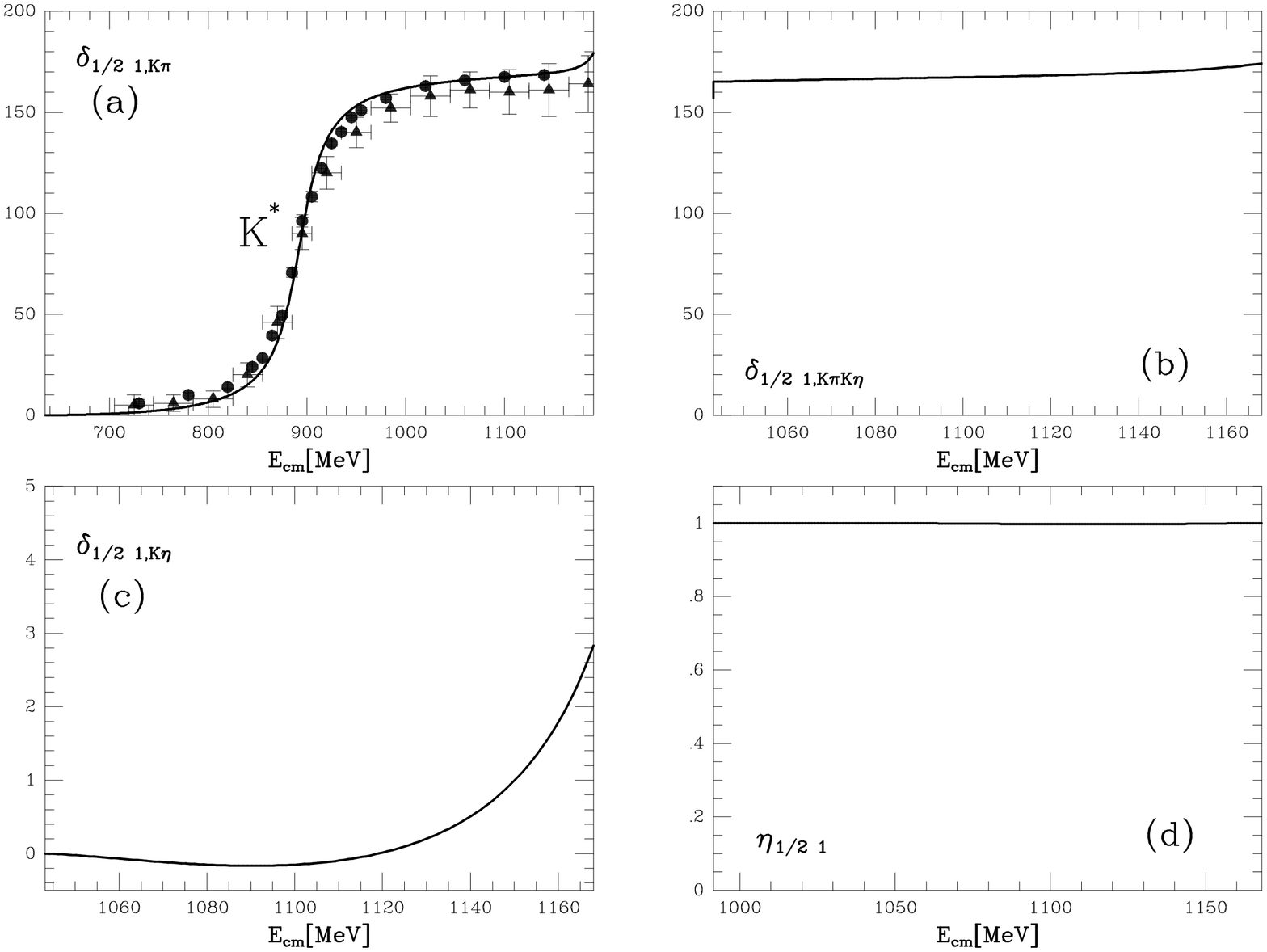,width=14cm,angle=-90}}
  {\footnotesize  Fig. 6: Results in the $I=1/2$, $J=1$ channel. (a) phase 
  shifts for $K\pi\rightarrow K\pi$. Data: full triangle \cite{MeAn71}, 
  open circle \cite{EsCa78}.  (b) phase shifts for
  $K\pi\rightarrow K\eta$. (c) phase shifts for $K\eta\rightarrow
  K\eta$. (d) inelasticity.}
\end{figure}

\begin{figure}
    \hbox{ \psfig{file=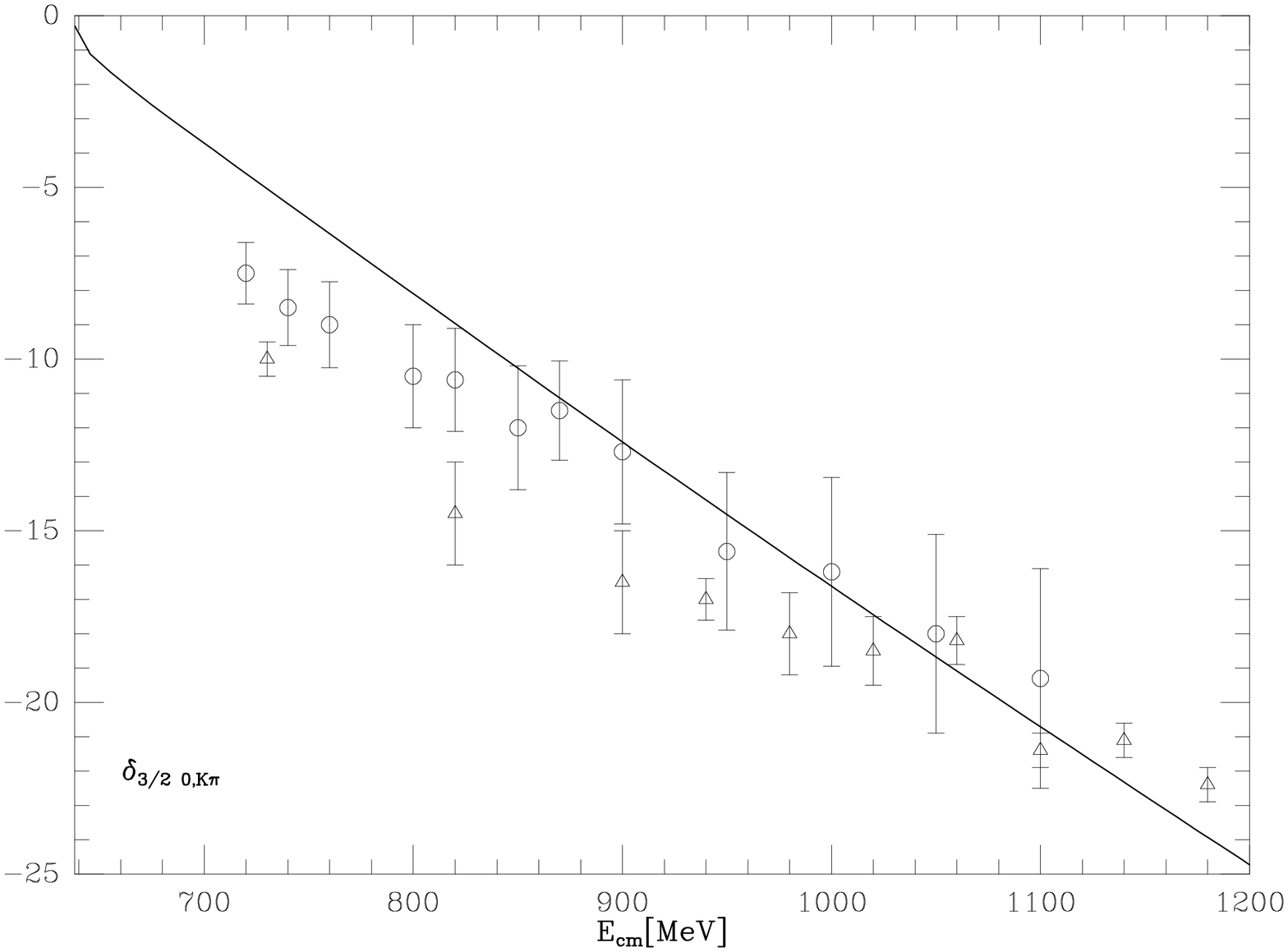,width=7cm,angle=-90}}
  {\footnotesize  Fig. 7: Phase shifts for $K\pi\rightarrow K\pi$ in the $I=3/2$,
  $J=0$ channel. Data: open triangle \cite{EsCa78}, open circle \cite{LiCh73}.}
\end{figure}

\begin{figure}
    \hbox{ \psfig{file=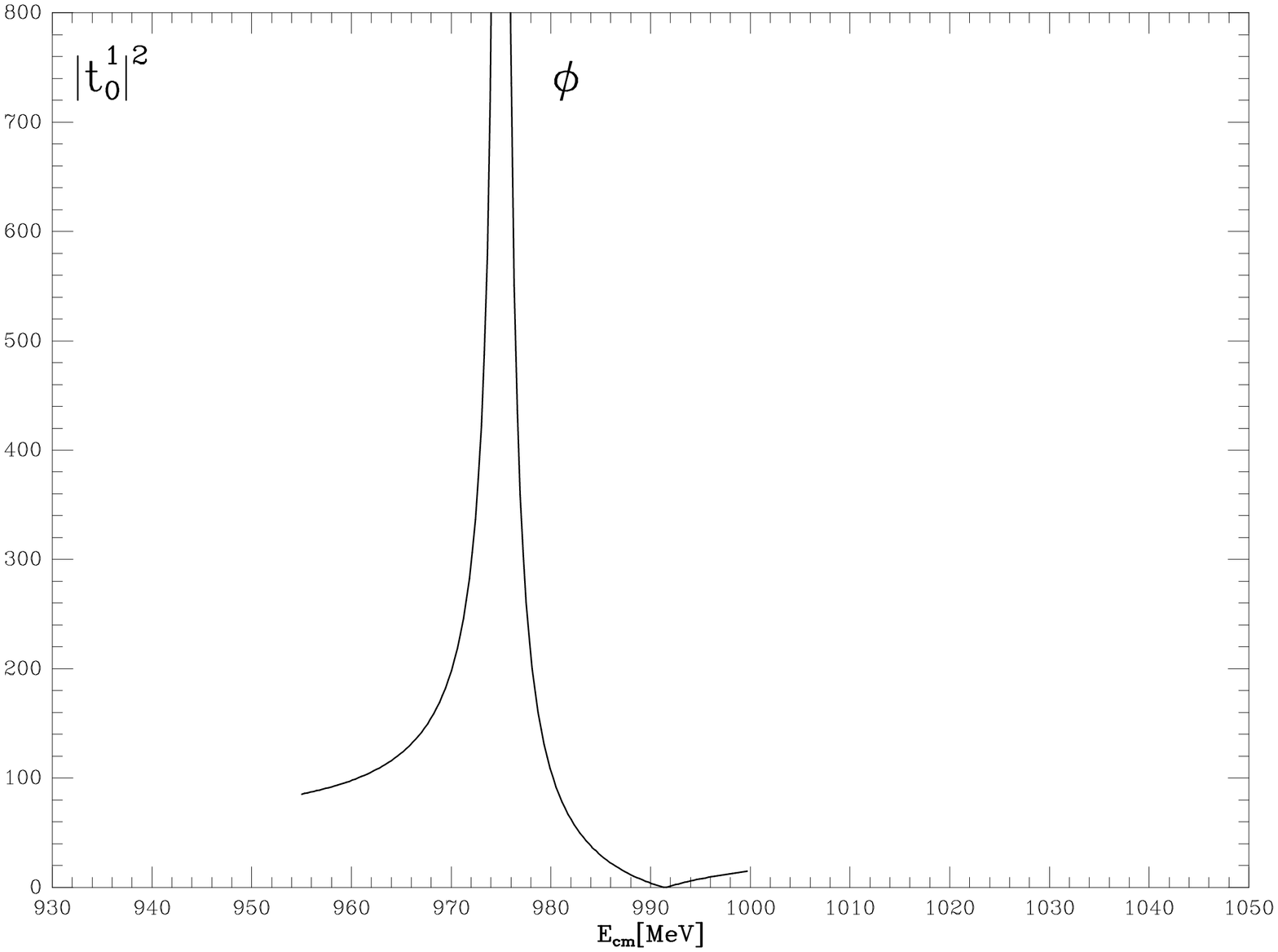,width=7cm,angle=-90}}
{\footnotesize Fig. 8: $(\vert T_{IJ=01} \vert)^2$ for $K \bar{K} \rightarrow K
  \bar{K}$ showing the singularity corresponding to the $\phi$
  resonance.
}
\end{figure}

\begin{figure}
\hbox{
\psfig{file=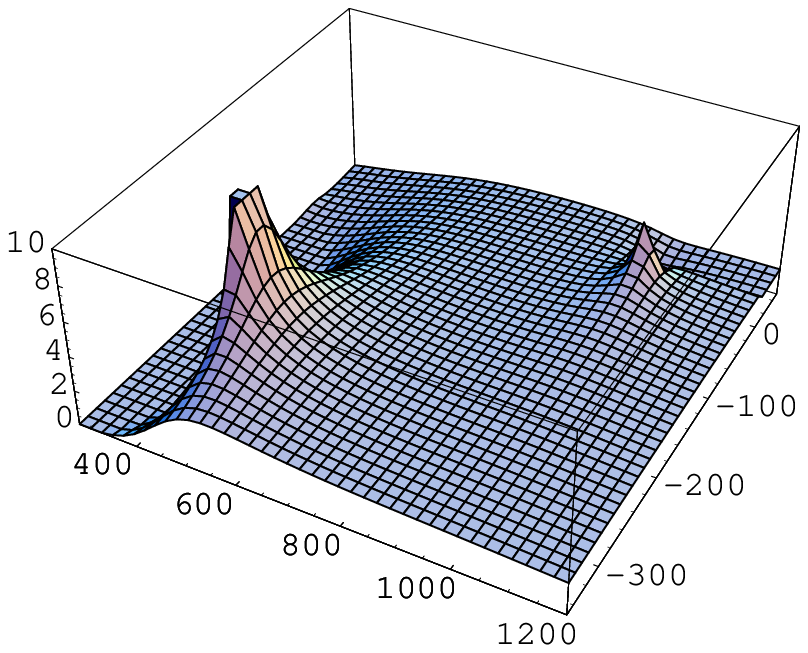,width=8cm}
\psfig{file=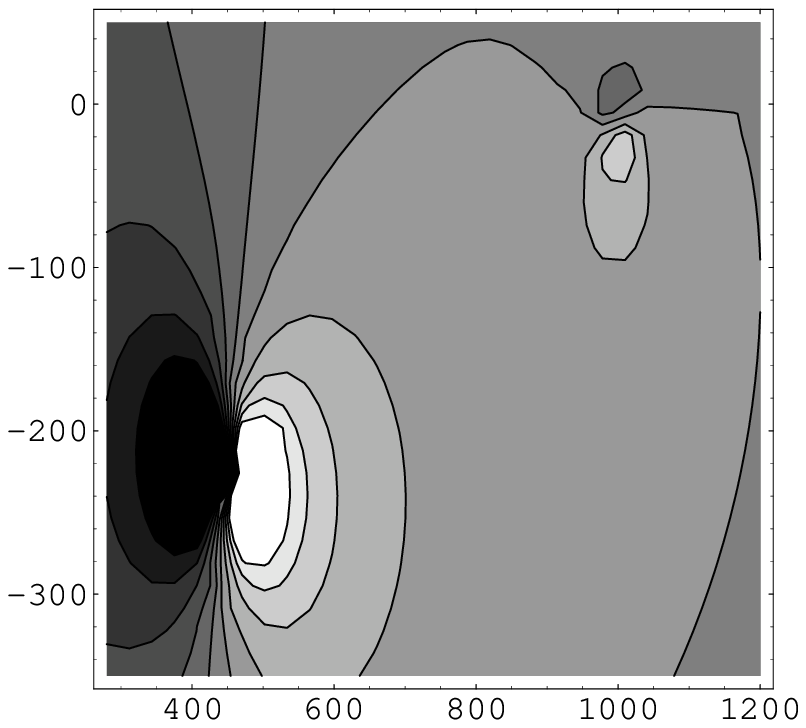,width=7.5cm}}
  {\footnotesize Fig 9: Imaginary part of the $\pi\pi$
  amplitude in the (I,J)=(0,0) channel in the second Riemann sheet.
On the left we show a three dimensional plot were we can observe the
different structure of the $\sigma$ and $f_0$ poles. On the right we
show a contour plot of the lower half plane of the second sheet. The
$\sigma$ pole is very far away from the real (physical) axis and its
lines of maximum gradient are parallel to it, in contrast with the
$f_0$. That is why the effect of both poles  in the phase shifts
(Figure 1) is so different.}
\end{figure}

\begin{figure}
\hbox{
\psfig{file=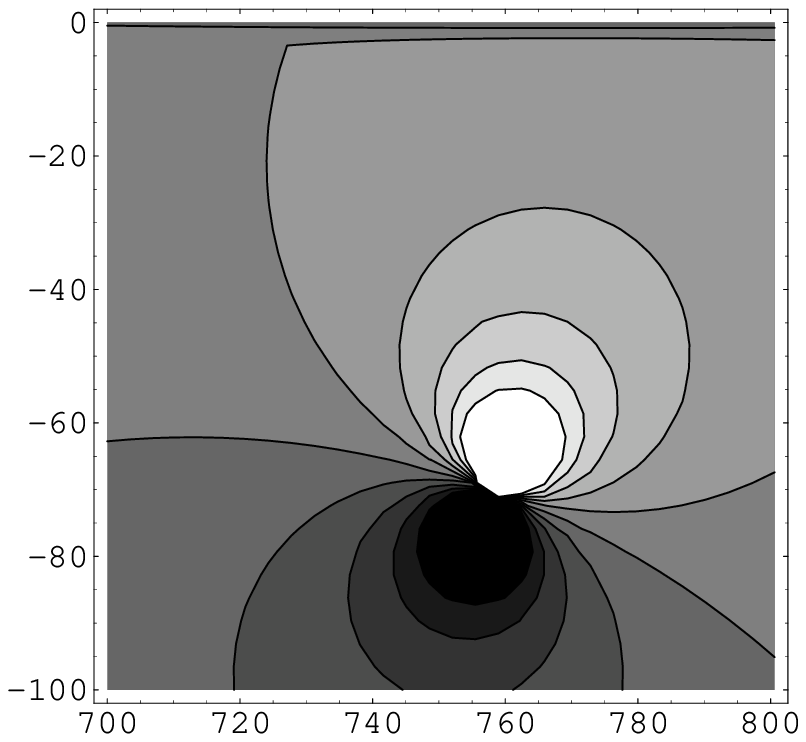,width=7.5cm}
\psfig{file=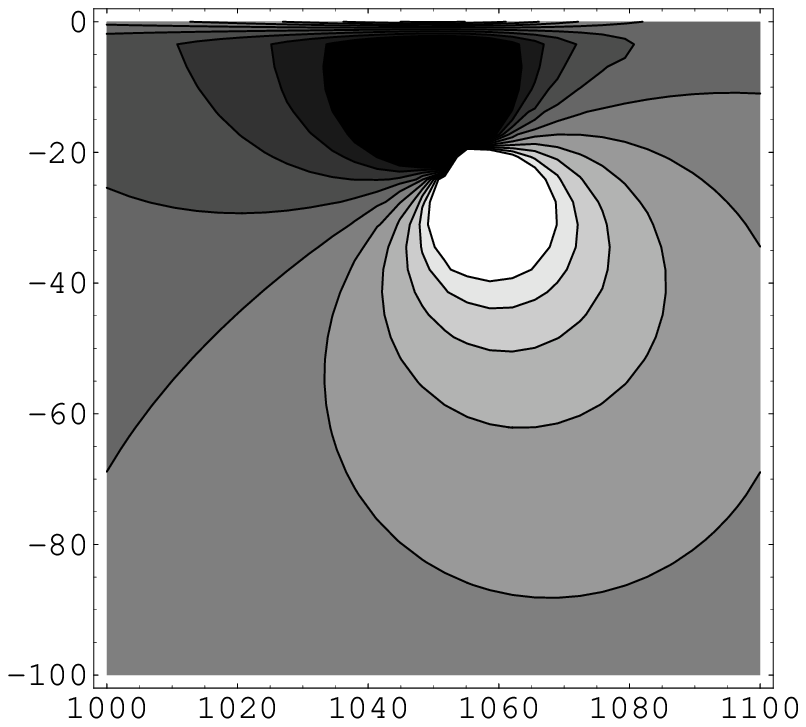,width=7.5cm}}
  {\footnotesize Fig 10: The poles associated to the $\rho$ (left) and $a_0$
  (right)  are oriented
  differently. The $\rho$ mass seen on the (I,J)=(1,1)
  phase shifts is slightly bigger than the real part of the position
  of the $\rho$ pole, whereas the peak of the mass distribution where
  the $a_0$ is observed (see Figure 4) is smaller than the real part of the $a_0$
  pole. Concerning the widths, they are obtained as twice the imaginary
  part of the associated pole position.}
\end{figure}

\newpage

\end{document}